# An All-Atom Force Field for Dry and Water-Lubricated Carbon Tribological Interfaces

Thomas Reichenbach[1,†], Severin Sylla[1,2,†], Leonhard Mayrhofer[1], Pedro Antonio Romero[3], Paul Schwarz[3], Michael Moseler[1,2,4,5], Gianpietro Moras[1,*]

[1]*Fraunhofer IWM, MikroTribologie Centrum µTC, Wöhlerstraße 11, 79108 Freiburg, Germany*

[2]*Institute of Physics, University of Freiburg, Hermann-Herder-Straße 3, 79104 Freiburg, Germany*

[3]*Freudenberg Technology Innovation SE & Co. KG, Hoehnerweg 2-4, 69469 Weinheim, Germany*

[4]*Freiburg Materials Research Center, University of Freiburg, Stefan-Meier-Straße 21, 79104 Freiburg, Germany*

[5]*Cluster of Excellence livMatS @ FIT – Freiburg Center for Interactive Materials and Bioinspired Technologies, University of Freiburg, Georges-Köhler-Allee 105, 79110 Freiburg, Germany*

*Corresponding author: gianpietro.moras@iwm.fraunhofer.de

[†] Thomas Reichenbach and Severin Sylla contributed equally.

**Abstract**

We present a non-reactive force field for molecular dynamics simulations of interfaces between passivated amorphous surfaces and their interaction with water. The force field enables large-scale dynamic simulations of dry and lubricated tribological contacts and is tailored to surfaces with hydrogen, hydroxyl and aromatic surface passivation. To favour its compatibility with existing force-field parameterizations for liquids and allow a straightforward extension to other types of surface passivation species, we adopt the commonly used OPLS functional form. The optimisation of the force-field parameters is systematic and follows a protocol that can be reused for other surface-molecule combinations. Reference data are calculated with gradient- and dispersion-corrected density functional theory and include the bonding structure and elastic deformation of bulk and surface structures as well as surface adhesion and water adsorption energy landscapes. The conventions adopted to define the different force-field atom types are based on the hybridisation of carbon orbitals and enable a simple and efficient parameter optimisation strategy based on quantum-mechanical calculations performed only on crystalline reference structures. Transferability tests on amorphous interfaces demonstrate the effectiveness of this approach. After testing the force field, we present two examples of application to tribological problems. Namely, we investigate relationships between dry friction and the corrugation of the contact potential energy surface and the dependency of friction on the thickness of interface water films. We finally discuss the limitations of the force field and propose strategies for its improvement and extension.

I. Introduction

Interfaces between amorphous carbon (a-C) surfaces play a particularly important role in the search of technological solutions to reduce friction and wear. This is mainly because of their ability to protect and passivate tribological surfaces in dry and boundary lubrication (BL) conditions. In BL, which often occurs in lubricated tribological systems at low sliding speed, high load or when using low-viscosity lubricants[1,2], surface asperities are separated at most by lubricant nanolayers and can come into direct contact with each other, typically causing high friction and wear. Not only are a-C surfaces able to mitigate friction and wear under these critical conditions, but they can even promote superlubricity (i.e., friction coefficient lower than 0.01) if they are used in the appropriate environment and if their chemical structure is properly tuned. This is for example the case of diamond coatings – whose surface structure is often amorphous[3] – in water vapor[4], tetrahedral amorphous carbon (ta-C) coatings in combination with glycerol or organic friction modifiers[5,6] or hydrogen-containing amorphous carbon (a-C:H) coatings in dry conditions[7–9]. Examples of a-C interfaces that can promote superlubricity do not only include diamond or diamond-like carbon coatings. An expanding area of research are self-generating a-C films that form under tribological load on ceramic or metallic surfaces, e.g., by tribochemical decomposition of C-containing molecules[10,11] or by material transfer from a carbon countersurface[8]. In general, achieving these levels of friction in BL requires control of the tribological interfaces down to the atomic level. For this reason, atomistic simulations have become indispensable to complement experiments and provide access to the atomic processes that determine friction.

Several studies investigated in detail the mechanisms underlying the BL properties of a-C interfaces, usually by combining tribometry, surface analytics and quantum-mechanical (QM) simulations[6,12–14]. A result that emerges from all these studies is that the reduction of friction and wear relies on a stable chemical passivation of the a-C surfaces as well as on their topography and the a-C's mechanical properties. Surface passivation prevents the formation of strong chemical bonds across the sliding interface, thus lowering the resistance to sliding. Depending on the kind of carbon film and the sliding environment (e.g., ta-C with organic friction modifiers or dry a-C:H), surface passivation can result from the tribochemical fragmentation of lubricants (e.g., water[12,15–17], glycerol or fatty acids[6]) and/or from shear-induced plastic deformation processes that take place at solid-solid contacts during the initial running-in stage[6,13,15,18]. After this initial high-friction phase, the friction coefficient drops as the surface dangling bonds are stably passivated by chemical species such as H and O atoms, OH groups[15–17,19], longer oligomers[13] or by the formation of aromatic surface structures[6,13,15,20].

All these kinds of surface passivation fulfil the task of preventing interface cold welding. However, friction coefficients can vary by about one order of magnitude (i.e., between ~0.01 and ~0.1) depending on the specific surface atomic structure (e.g., the surface density of passivation species, their length and chemical composition[21]) and on the presence or absence of residual lubricant nanofilms at contacts between asperities. For example, molecular dynamics (MD) simulations[21] show friction coefficients of about 0.2 for dry contacts between a-C surfaces passivated by chemisorbed hydrocarbon chains composed of 5 to 10 C atoms. This friction coefficient is about one order of magnitude higher than in the case in which the same surface is terminated by H atoms. This difference is related to the steric interdigitation of the surface chains and depends mainly on their surface density. A quantitatively similar difference was also observed by means of QM simulations in the completely different case of water-lubricated diamond[15]. In this example, the presence of 1 to 3 water monolayers between OH/H-terminated surfaces is characterised by friction coefficients of about 0.2, while in absence of residual water molecules at the passivated interface the friction coefficients are about 0.02. In this case, the precise mechanisms responsible for the difference in friction are not fully understood. These two examples show how achieving superlubricity in BL requires an optimal structure of the sliding interface. The latter should minimise the energy barriers for sliding (i.e., the corrugation of the contact potential energy surface, CPES)[22] and ensure a high stiffness of the tribological contact[23,24]. Atomistic simulation studies show that this can be achieved when the surfaces of contacting a-C asperities are atomically smooth, passivated by short terminations (e.g. H, O) or aromatic structures and there are no residual lubricant nanolayers between them[6,15,21]. This is conceptually similar to the case of structural superlubricity between flat, incommensurate crystalline surfaces[25,26].

This scenario illustrates how in BL, unlike in hydrodynamic lubrication, the atomic-scale structure of the a-C interfaces has a decisive effect on friction. Besides governing the interaction between surfaces under dry conditions, the surface structure also strongly influences the slip length[27] and the structure of lubricating films for distances up to several nanometers from the surface. This is particularly relevant if the total film thickness falls in this order of magnitude[28–30]. The design of optimal surface-lubricant combinations for boundary lubricated systems would benefit considerably from the knowledge of structure-property relationships that connect the interface atomic structure with friction. These relationships can only be obtained systematically by combining experiments with atomistic simulations, which provide access to the atomic-scale details of the sliding interface. However, the computational cost of QM simulation methods that can accurately describe chemically complex systems is currently too high for this kind of studies and more efficient reactive interatomic potentials for C, O, H systems are not available yet[31,32] or are not tailored for tribological systems[33]. A possible way forward is the use of non-reactive force fields (FFs), like those routinely used for liquids[34] and often employed for the study of lubricants' rheology[35,36]. These FFs enable a computationally efficient treatment of systems with many chemical elements, they are well tested on a large number of molecules that are used in lubricants and on functional groups that can be found on a-C surface passivation layers. Yet, they are rarely developed for solids and their surfaces.

Therefore, a specific parametrization of a FF for a-C materials and their elastic properties, which are important in BL[23], needs to be developed from scratch, which is one of the main goals of this work. A very useful feature of non-reactive FFs is their very simple functional form whose parameters can be individually modified, for example to study the effect of single atomic-scale features (e.g., bond stiffness or atomic charge) on friction[24,37]. Their major limitation, however, is that bonds can neither form nor break during a MD simulation, i.e., FFs cannot describe chemical reactions and rely on predefined system topologies, usually derived by accessory reactive simulations[21,24]. For this reason, they have to be used with care and in situations in which the process of interest is dominated by physical interactions rather than by chemical reactions, as is often the case in the simulation of lubricants' rheological properties. In the specific applications targeted in this work, the use of a non-reactive FF can certainly be justified during the low-friction steady-state that follows the initial running-in phase. QM MD simulations show that in this situation passivated a-C interfaces do not experience chemical reactions for the whole duration of ns-long simulations, even under high normal pressures[6,13,15]. Moreover, ultralow or superlow friction coefficients in experiments are stable for time scales that are orders of magnitude longer than those accessed by MD simulations[4,6,7].

MD simulations based on non-reactive FFs have been used in many studies of tribological interfaces in BL, often in cases where the importance of the solid elastic response is negligible owing to the presence of soft chemisorbed layers in which most of the shear deformation is localised. Examples include the study of friction as a function of parameters like load, sliding speed, lubricant composition and atomic-scale features of the surface for self-assembled monolayers on silica[38], hexadecane nanofilms confined between iron oxide surfaces[39], organic friction modifiers adsorbed on iron oxide and lubricated by hexadecane[36], interfaces between surfactants and iron oxide[40]. With regard to tribological systems with carbon surfaces, we developed bespoke versions of the OPLS ("Optimized Potential for Liquid Simulations") FF[34] to study the relationship between friction, CPES corrugation, and interdigitation of surface passivation species in dry contacts between hydrogenated and fluorinated diamond surfaces[24], and between a-C surfaces passivated by chemical species with different length and polarity[21].

The goal of the present study is to extend and generalise our previous work[21,24,41] by developing a non-reactive FF that is compatible with existing parameterizations of the OPLS FF for liquids and enables simulations of both dry and lubricated contacts between passivated a-C surfaces, with particular focus on their relative motion under a normal load. To favour compatibility with existing OPLS parametrizations for lubricants, we use the OPLS functional form and only develop parameters for bulk a-C and its passivated surfaces, without altering the original parameters for the lubricant molecules. Moreover, to model surface terminations, we use functional groups for which OPLS parameters already exist and we test these parameters before further optimizing them when needed. To limit complexity in this initial development stage, here we focus only on monovalent H and OH surface terminations as well as aromatic surface structures, and use water as a model lubricant, also because of its importance in achieving superlubricity in various systems[4,42–44]. However, given our development strategy, extending the FF to other surface terminations and lubricating molecules should be relatively straightforward, as we will show in the final part of the article.

Optimisation of existing parameters is not possible for bulk a-C for which no FF is available. Yet the simulation of asperity contacts requires a correct description of the elastic response of bulk a-C and of its terminations. This is because friction depends on the system's stiffness in addition to the CPES[23,24] as shown for instance by the Prandtl-Tomlinson model of dry, wearless friction[23]. We therefore develop the FF for a-C from scratch. In this phase, we aim to be as general as possible and systematically derive a FF that is applicable over a wide range of a-C densities[45,46], from 1.75 g/cm$^3$ to 3.5 g/cm$^3$. This aspect is crucial, as the a-C density determines its elastic properties but also its surface structure, with consequences on the atomic-scale surface corrugation and on the density and type of surface termination groups.

All reference data used to fit and test the FF are calculated with gradient- and dispersion-corrected density-functional theory (DFT)[47–50]. The way the various types of atoms are defined in the FF is mainly based on the coordination of the carbon atoms. This allows an optimisation of the corresponding parameters on well-defined crystalline reference structures (i.e., diamond, graphite and graphene structures). In addition to being simple, this procedure significantly reduces the computational cost of fitting as DFT simulations do not have to be performed on a large number of amorphous structures, which would be necessary to obtain a representative sampling. The reference data used for optimizing and testing the FF parameters include the electrostatic potential (ESP) above the surfaces, the elastic response of surface groups and the bulk material, the surface adhesion, and the adsorption of lubricant molecules. The effectiveness of this approach, which is based on the similarity between local atomic environments in crystalline and amorphous carbon, is demonstrated by transferability tests that were carried out on amorphous interfaces and, besides the quantities listed above, also include contact potential energy (CPE) profiles. An example of how to setup FF simulations with the LAMMPS[51] code is provided in the Python package matscipy[52].

The article is organised as follows. Section II describes the functional form of the FF and introduces the different atom types used to describe crystalline and amorphous bulk carbon structures and surfaces. In Section III, we present the optimisation of the FF parameters based on crystalline reference systems. This is followed in Section IV by accuracy and transferability tests of the optimised parameter set on crystalline and amorphous systems. Section V deals with two example applications of the FF to tribological problems: the relationship between dry friction and CPE corrugation, and the relationship between friction and water film thickness. Finally, in Section VI, we discuss possible future extensions and improvements of the FF.

## II. Force field

The goal of this work is the development of a non-reactive FF for MD simulations of dry and water-lubricated tribological interfaces between a-C surfaces (example configurations in Figure 1a). This implies an accurate description of the surface-surface, surface-water and water-water physical interactions as well as of the elastic response of the system. The FF should be able to do so for a wide range of a-C densities (1.75 – 3.5 g/cm$^3$), spanning form low-density, sp$^2$-rich a-C to sp$^3$-rich, diamond-like a-C (e.g., tetrahedral amorphous carbon)[45,46]. Of course, since the FF is non-reactive, a-C systems must be prepared beforehand using, e.g., reactive MD simulations, as described in Section IV.2. In general, the dangling bonds on a-C surfaces in contact with water can be passivated by H atoms, OH groups or O atoms in a variety of configurations (e.g., ether and carbonyl groups)[4,12,15–17]. To limit the initial complexity of the model, we here focus on surface passivation by H atoms and OH groups only. Finally, since localised aromatic structures on a-C surfaces were proposed to play a relevant role in surface passivation and friction of a-C tribological systems[6,13,15,20], we also consider their description when developing the FF.

In order to easily combine the FF for a-C bulk and surfaces described here with existing and well-tested FFs for liquids, including water, we choose the OPLS functional form[34]. The latter is described in Section II.1 so that the reader can more easily follow the description of the FF parameterisation. Within this non-reactive framework, a set of atom types, into which each atom can be uniquely categorised, must be defined to account for different chemical environments. The atom types introduced for this FF are schematically summarised in Figure 1b. They are defined in Table 1 and discussed in detail in Section II.2 for bulk a-C and in Section II.3 for the surfaces. The interactions between atom types are defined by the OPLS functional form, which includes bonded interactions (bond stretching, angle bending and dihedral rotations) and non-bonding interactions (Lennard-Jones and Coulomb potentials). The active potential energy terms for the different atom types in bulk a-C, a-C surfaces and water, respectively, are summarised in Figure 1c.

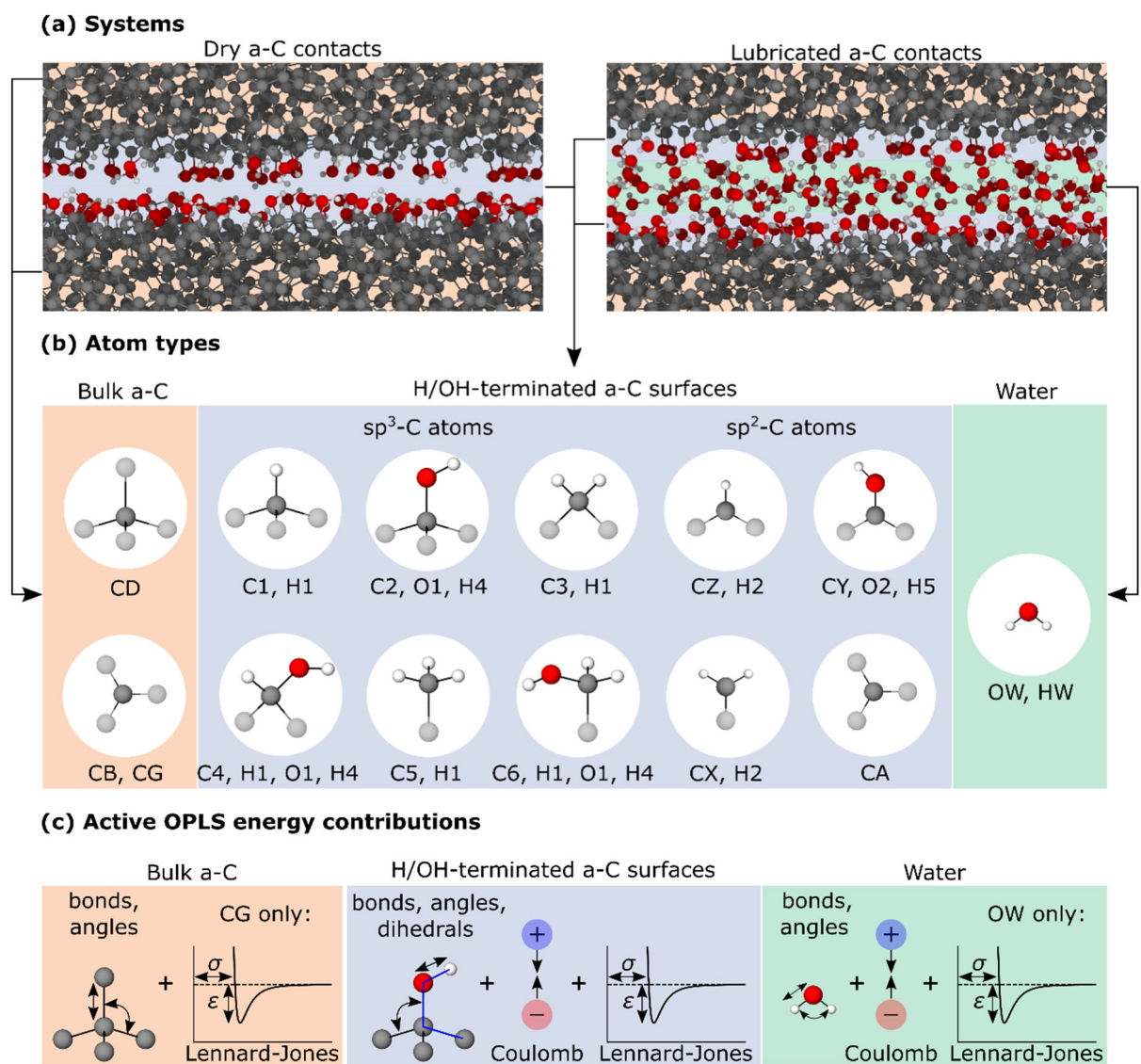

*Figure 1 (a) Example configurations of dry and water-lubricated a-C interfaces. Carbon, oxygen and hydrogen atoms are coloured grey, red and white, respectively. The bulk, surface and lubricant regions are indicated in orange, blue and green, respectively. (b) Overview of the different atom types and (c) schematic illustration of the different potential energy contributions used to describe the bulk, surface, and lubricant regions (see text for details). The labels used for each atom type throughout this article are reported below each configuration in panel (b). Light grey carbon atoms in panel (b) are first neighbours of the carbon atoms defined by the labels.*

*Table 1. Overview of the atom types used in the FF for the description of OH/H-terminated carbon surfaces in dry and humid conditions. The first section corresponds to bulk atom types, the second section to surface atom types and the third section to the TIP3P water model that is adapted from Ref. 53. See also Figure 1b for a schematic visualization of the atom types.*

| Atom type | Description |
|---|---|
| CD | $sp^3$-C atom with 4 C neighbours |
| CB | Bulk sp- or $sp^2$-C atom with 2 or 3 C neighbours, respectively; at least one neighbour is a $sp^3$-C atom. |
| CG | Bulk sp- or $sp^2$-C atom with 2 or 3 C neighbours, respectively; no $sp^3$-C neighbours. |
| C1 | $sp^3$-C atom with 3 C neighbours and 1 H neighbour |
| C2 | $sp^3$-C atom with 3 C neighbours and 1 OH group |
| C3 | $sp^3$-C atom with 2 C neighbours and 2 H neighbours |
| C4 | $sp^3$-C atom with 2 C neighbours, 1 H neighbour and 1 OH group |
| C5 | $sp^3$-C atom with 1 C neighbour and 3 H neighbours |
| C6 | $sp^3$-C atom with 1 C neighbour, 2 H neighbour and 1 OH group |
| CZ | $sp^2$-C atom with 2 C neighbours and 1 H neighbour |
| CY | $sp^2$-C atom with 2 C neighbours and 1 OH group |
| CX | $sp^2$-C atom with 1 C neighbour and 2 H neighbours |
| CA | C in graphite or $sp^2$-CB/CG surface atom in a-C |
| O1 | Oxygen atom of OH termination on $sp^3$-C |
| O2 | Oxygen atom of OH termination on $sp^2$-C |
| H1 | H termination on $sp^3$-C |
| H2 | H termination on $sp^2$-C |
| H4 | H atom of OH termination on $sp^3$-C |
| H5 | H atom of OH termination on $sp^2$-C |
| OW | O atom of TIP3P water molecules, parameters from Ref. 53 |
| HW | H atom of TIP3P water molecules, parameters from Ref. 53 |

## II.1 Functional form

The potential energy for a configuration of $N$ atoms with positions $\vec{r}_1, \dots, \vec{r}_N$ in OPLS reads

$$E(\vec{r}_1, \ldots, \vec{r}_N) = E_{\text{bonded}}(\vec{r}_1, \ldots, \vec{r}_N) + E_{\text{nonbonded}}(\vec{r}_1, \ldots, \vec{r}_N).$$

(1)

$E_{\text{bonded}}$ describes intramolecular interactions due to covalent bonds and is composed of harmonic bond-stretching $E_{\text{bond}}$ and angle-bending $E_{\text{angle}}$ terms, and a dihedral term $E_{\text{dihedral}}$

$$E_{\text{bonded}} = E_{\text{bond}} + E_{\text{angle}} + E_{\text{dihedral}},$$

(2)

where

$$E_{\text{bond}} = \sum_{\alpha \in \{\text{bonds}\}} k_{r,\alpha}(r_\alpha - r_{\text{eq},\alpha})^2,$$

(3)

$$E_{\text{angle}} = \sum_{\beta \in \{\text{angles}\}} k_{\theta,\beta}(\theta_\beta - \theta_{\text{eq},\beta})^2,$$

(4)

$$E_{\text{dihedral}} = \sum_{\gamma \in \{\text{dihedrals}\}} \frac{k_{\Phi,\gamma,1}}{2}(1 + \cos(\Phi_\gamma)) + \frac{k_{\Phi,\gamma,2}}{2}(1 - \cos(2\Phi_\gamma)) + \frac{k_{\Phi,\gamma,3}}{2}(1 + \cos(3\Phi_\gamma)).$$

(5)

Here, $r_\alpha$ refers to the length of bond $\alpha$ with equilibrium bond length $r_{\text{eq},\alpha}$, $\theta_\beta$ denotes the angle $\beta$ with equilibrium value $\theta_{\text{eq},\beta}$, and $\Phi_\gamma$ is the dihedral angle $\gamma$. $k_r$ and $k_\theta$ are two times the force constants for bond stretching and angle bending, and $k_{\Phi,\gamma,1}, k_{\Phi,\gamma,2}, k_{\Phi,\gamma,3}$ are Fourier coefficients.

The nonbonded energy term is the sum of a Coulomb and a Lennard-Jones (LJ) contribution: $E_{\text{nonbonded}} = E_{\text{Coulomb}} + E_{\text{LJ}}$. Electrostatic interactions are modelled via

$$E_{\text{Coulomb}} = \sum_i \sum_{j>i} \frac{q_i q_j e^2}{r_{ij}},$$

(6)

where $r_{ij}$ is the distance between the atoms $i$ and $j$ with charges $q_i$ and $q_j$ that are measured in units of the elementary charge $e$. The short-range Pauli repulsion and the long-range dispersion interactions are represented by

$$E_{\text{LJ}} = \sum_i \sum_{j>i} 4\varepsilon_{ij} \left( \left(\frac{\sigma_{ij}}{r_{ij}}\right)^{12} - \left(\frac{\sigma_{ij}}{r_{ij}}\right)^6 \right),$$

(7)

where $\varepsilon_{ij}$ and $\sigma_{ij}$ are the energy minimum of the pairwise interaction and the interatomic distance at which the LJ potential is zero, respectively. The two-body parameters $\varepsilon_{ij}$ and $\sigma_{ij}$ are obtained from the atomic parameters $\varepsilon_i$ and $\sigma_i$ via $\varepsilon_{ij} = \sqrt{\varepsilon_i \varepsilon_j}$ and $\sigma_{ij} = \sqrt{\sigma_i \sigma_j}$ [34]. We truncate and shift the LJ interaction to zero at a distance of 14 Å, while we use a particle-particle particle-mesh solver[54] for the electrostatic interactions, which makes them effectively infinite-ranged. Nonbonded interactions between atom pairs that are separated by less than three bonds are omitted, and for atom pairs that are separated by exactly three bonds the nonbonded interaction is halved[34]. All FF calculations are carried out with LAMMPS[51].

## II.2 Description of bulk diamond, graphite, and amorphous carbon

The first, critical step of the FF development is the choice of the atom types that are used to categorise the atoms of the simulation domain. While in quantum-mechanical methods and reactive interatomic potentials the chemical element uniquely identifies each atom, several atom types for the same chemical element can be required in non-reactive FFs to account for different chemical environments. For example, a hydrogen atom bound to a carbon atom will have a different atom type (e.g., H1 in Figure 1b) than a hydrogen atom in a hydroxyl group (e.g., H4 in Figure 1b). A description of all atom types of our FF is provided in Table 1 and in Figure 1b.

The definition of the atom types can become an extremely complex task when applied to amorphous carbon, in which each carbon atom can have significantly different chemical environments. To simplify this task and the FF itself, we adopt and further develop a strategy we proposed in Ref. 24. We think of a-C as composed of two kinds of carbon atoms: C with $sp^3$ hybridization ($sp^3$-C) and C with $sp^2$ hybridization ($sp^2$-C). The FF parameters for these two atom types can be conveniently obtained by reproducing the relevant properties of diamond and graphite crystals as calculated with the reference DFT method.

To describe bulk diamond we follow the approach we introduced in Refs. 24,41, where we showed that a simple FF consisting solely of harmonic bonds and angles can accurately describe the geometry and the elastic constants of diamond. The corresponding atom type is labelled CD, has four carbon nearest neighbours (Figure 1b), zero charge ($q_{CD} = 0$) and does not interact with other atoms via LJ potentials (i.e., $\varepsilon_{CD} = 0$). Similarly, to describe graphite, we define the atom type CA that has three carbon nearest neighbours and zero charge. In this case, in addition to harmonic bonds and angles, LJ interactions are required to describe the interlayer spacing and the elastic properties in the direction perpendicular to the graphite layers.

As we will show in Section IV.2, the structure and elastic properties of a-C can be accurately described simply using CD and CA with some minor modifications. Carbon atoms with 4 C neighbours (here determined by using a cutoff distance of 1.85 Å[46]) are classified as CD. While the CA atom type is suitable for describing $sp^2$-C atoms on a-C surfaces, simply assigning the CA type to $sp^2$-C atoms in bulk a-C leads to unphysical attractive LJ interactions between $sp^2$-C that are far apart and separated by other C atoms that should screen such interactions. This leads to an overestimation of the elastic moduli for high density a-C. At the same time, omitting the LJ interactions leads to an overestimation of the density for low-density a-C with localised graphitic structures[46], where a repulsion between adjacent graphitic layers is required. As a remedy, we derive two additional atom types from CA to describe C atoms with 3 C neighbours in a-C. These two atom types are labelled CB and CG. The type CB is used for $sp^2$-C atoms that are not embedded in a local graphitic structure (i.e., at least one neighbour is a $sp^3$-C atom). CB has the same parameters as CA but does not interact with other atoms via LJ interactions ($\varepsilon_{CB} = 0$). The CG atom type is used for C atoms with no $sp^3$-C neighbours. CG has the same LJ parameters as CA, but the CG–CG and CG–CA LJ interactions are purely repulsive, i.e., their LJ interaction is truncated at the energy minimum where the potential is shifted to zero. In accordance with Ref. 24, we also use the CG and CB parameters for bulk sp-hybridised C atoms, which appear in relatively small quantities. The parameters for harmonic bonds between $sp/sp^2$-C and $sp^3$-C atoms are chosen to be arithmetic averages of the parameters for CD–CD and CA–CA bonds. For C–C–C angles that are centred on a $sp/sp^2$-C atom, we use the CA–CA–CA parameters. Analogously, we use the CD–CD–CD parameters for C–C–C angles centred on $sp^3$-C atoms.

II.3 Description of carbon surfaces and water

To model the functional groups on H-/OH-passivated a-C surfaces, we adopt the atom types and conventions used in the OPLS FF for hydrocarbons and alcohols[34]. Similarly to what was described previously for the atom types in bulk a-C, we consider $sp^3$- and $sp^2$-C surface atoms, whose dangling bonds are passivated by H atoms or OH groups (see Figure 1b and the description in Table 1). For simplicity, and following the conventions used in OPLS for alcohols[34] we consider at most one OH group per surface carbon. The atom types C1, C3 and C5 are $sp^3$-C atoms bonded to 1, 2 or 3 H atoms (atom type H1), respectively (all other neighbours are C atoms). The atom types C2, C4 and C6 are simply obtained from C1, C3 and C5 by replacing one H atom with an OH group (atom types O1 and H4). In a similar way, we define four atom types for surface $sp^2$-C atoms: CZ has two C and one H neighbours (H2, like H in benzene); CY is similar to CZ but H is replaced by OH (O2, H5, like OH in phenol); CX has one $sp^2$-C and additionally two H neighbours (H2, like H in a methylidene group); CA is a surface $sp^2$-C atom with three C neighbours as in a surface aromatic structure[6] and has been already defined in Section II.2. The surface region to which CA-type atoms belong is arbitrarily defined here and extends 2 Å into the bulk from the outermost C atom.

For bonded interactions involving both surface and bulk C types, we treat the surface C types as their bulk counterparts, i.e., surface $sp^3$-C types, like C1, are treated like CD and surface $sp^2$-C types, like CZ, are treated like CG or CB. To account for the energy barriers for the rotation of surface O1–H4 and O2–H5 groups and to penalise the out-of-plane bending of CZ–H2 terminations, we introduce dihedral terms (Eq. ( 5 )) that are defined in Table S4. In these cases, we simply reuse the original OPLS dihedral parameters for alcohols, phenol and benzene[34,55] since we find that a further parameter optimisation is only beneficial in some isolated cases and can cause a loss of transferability. As in the a-C bulk case, we optimise the FF parameters using well-defined auxiliary crystalline cases and subsequently reuse or adapt the parameters for a-C surfaces. In particular, we use H-/OH-passivated diamond (111) surfaces and zigzag graphite edges as crystalline models for surface $sp^3$-C and $sp^2$-C atoms, respectively. Lastly, we use the well-established TIP3P model with long-range electrostatic interactions for the description of water[53,56], which is often used in conjunction with OPLS FFs. The corresponding atom types are labelled OW and HW in Figure 1 and Table 1.

### III. Parameter optimisation

Our parameter optimisation procedure consists of 6 steps as illustrated in Figure 2. In each step a different subset of FF parameters is optimised by fitting the appropriate reference data as calculated by DFT simulations (details are provided below). The ordering of the optimisation steps is chosen according to the mutual dependencies of the parameters (as indicated by arrows in Figure 2) to keep the optimisation problem as simple as possible. The six optimisation steps are described in detail in the next subsections III.1-III.6 and can be briefly summarised as follows:

1. The parameters for bonded interactions (bonds and angles) between bulk $sp^3$-C atom types (CD) are determined by fitting the lattice and elastic constants of diamond.

2. The LJ parameters for bulk and surface $sp^2$-C atom types in graphitic environments (CG, CA) are optimised by fitting the energy dependency on interlayer separation in graphite.

3. Parameters for bonded interactions (bond and angles) between $sp^2$-C bulk and surface atoms (CG, CB, CA) are determined by fitting the lattice and elastic constants of graphene (these quantities depend on the LJ parameters as well).

4. The atomic charges of the C, O and H surface types are determined by fitting the ESP near diamond surfaces and graphite edges and are therefore independent of the LJ and bonded interactions.

5. The LJ parameters of the C, O and H surface types are optimised by fitting energy-distance curves for the adhesion between diamond (111) surfaces and between zigzag graphite edges, as well as for the adsorption of (TIP3P) water molecules on these two crystalline reference systems.

6. The bonded parameters (bonds and angles) involving surface C, O and H atom types are optimised by fitting bond stretching and angle bending energies on the diamond surface and graphite edge reference systems. As mentioned in Section II.3 dihedral parameters are taken directly from OPLS whenever needed.

To make the surface parameter optimisation independent of the bulk bonding parameters, we use the DFT-optimised atomic configurations to evaluate the FF potential energies in the steps 4–6. A complete list of all optimised parameters is provided in Section S1 in the Supporting Information.

DFT simulations to generate all fit and test data are carried out with the Quickstep module of the CP2K[50] software package. We use the PBE exchange-correlation functional[49] in conjunction with the D3(BJ) van der Waals correction[47,48], GTH pseudopotentials[57] and the DZVP-MOLOPT-SR-GTH basis set[58]. We use a plane-wave cutoff of at least 15.0 keV. The Brillouin zone is sampled with the Γ-point for all amorphous structures and the crystalline structures discussed in the Sections III.1, III.3 and IV.1 (adhesion and adsorption with relaxed geometries), while we employ a Monkhorst-Pack sampling scheme[59] for the energies and forces of the crystalline structures discussed in the Sections III.2 (22 × 22 × 9 grid), III.4, III.5, and III.6 (a 7 × 7 × 1 grid for diamond and a 9 × 9 × 1 grid for graphite). Surfaces are modelled as slabs with periodic boundary conditions in all directions and a vacuum region of at least 20 Å between neighbouring images of slabs along the surface normal.

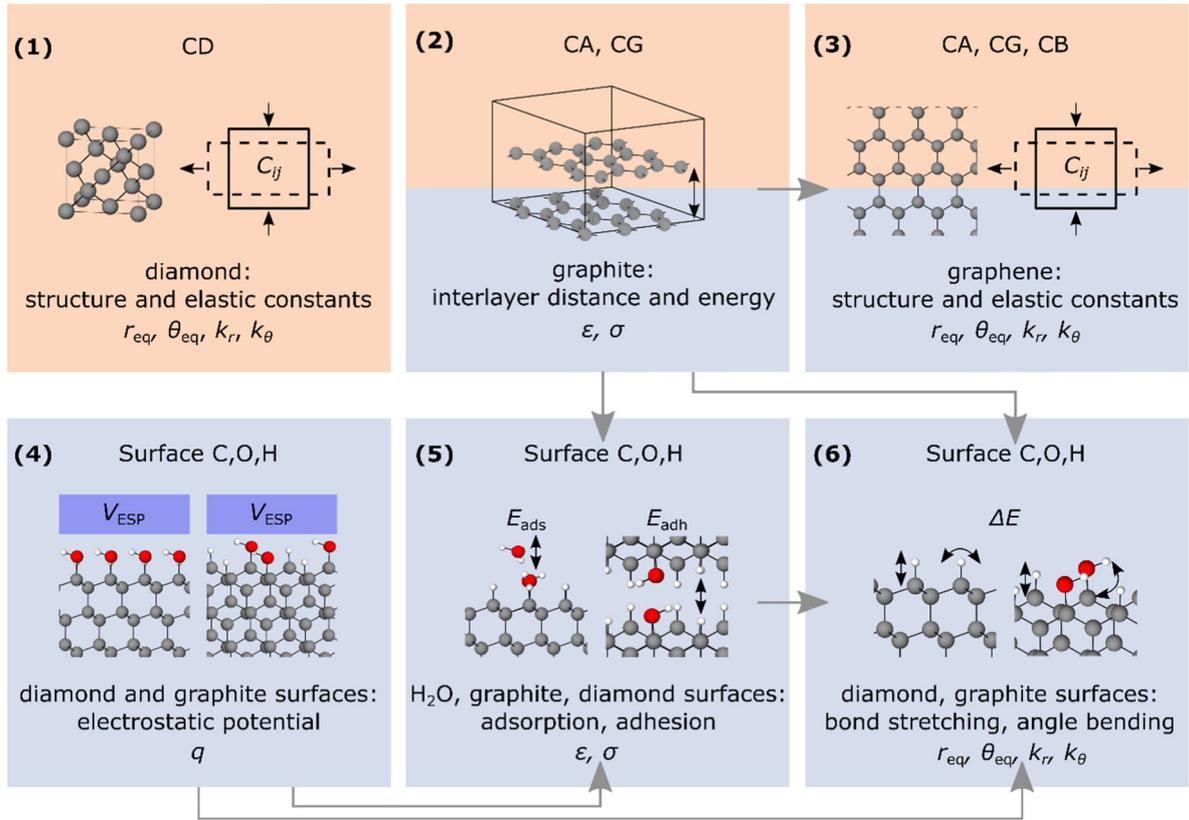

*Figure 2 Schematic overview of the parameter optimisation procedure. In each of the six steps the parameters listed at the bottom of each box are optimised using the DFT reference data schematically represented in the centre of the box. The relevant atom types are reported at the top of each box. The background colours indicate the kind of atom types that are optimised in each step: bulk atom types (orange) or surface atom types (blue). Grey arrows indicate dependencies of optimisation steps (e.g., the result of the optimisation in step (3) depends on the parameters optimised in step (2)).*

## III.1 Bonded interactions between sp³-C atom types

To optimise the parameters of the bonded interactions involving the CD atom type, we consider a periodic, cubic diamond supercell with 216 atoms. The DFT-optimised lattice constant of $a = 3.563$ Å directly yields $r_{eq,\text{CD-CD}} = \frac{\sqrt{3}}{4} a = 1.543$ Å, while the equilibrium angle $\theta_{eq,\text{CD-CD-CD}} = 109.5°$ is chosen according to diamond's tetrahedral structure. $k_{r,\text{CD-CD}}$ and $k_{\theta,\text{CD-CD-CD}}$ are tuned to fit the DFT values of diamond's elastic constants $C_{11}$, $C_{12}$ and $C_{44}$ (Table 2). To do so, we minimise the following objective function using SciPy[60]:

$$\Xi_{\text{bulk}}(k_r, k_\theta) = \sqrt{\sum_{(i,j)} \left( \frac{C_{ij,\text{FF}}(k_r, k_\theta) - C_{ij,\text{DFT}}}{C_{ij,\text{DFT}}} \right)^2 },$$

(8)

where $C_{ij,\text{FF}}$ are the elastic constants as obtained by FF calculations, $C_{ij,\text{DFT}}$ are the DFT elastic constants and $(i,j) \in \{(1,1), (1,2), (4,4)\}$.

*Table 2. Elastic constants of diamond and graphene as determined by DFT and by the present FF. The elastic constants are calculated by the slope of stress-strain curves when applying small triclinic simulation cell deformations using strains of up to 1%. The DFT elastic constants were calculated with matscipy[52] and the FF elastic constants with LAMMPS[51]. The elastic constants of graphene are given in N/m since graphene is a two-dimensional material.*

|  | DFT | FF |
|---|---|---|
| **Diamond** | | |
| $C_{11}$ (GPa) | 1067 | 1151 |
| $C_{12}$ (GPa) | 127 | 127 |
| $C_{44}$ (GPa) | 574 | 519 |
| **Graphene** | | |
| $C_{11}$ (N/m) | 365 | 353 |
| $C_{12}$ (N/m) | 60 | 61 |
| $C_{66}$ (N/m) | 143 | 146 |

### III.2 Lennard-Jones parameters of sp$^2$-C atom types

For the optimisation of the CA LJ parameters, we use interlayer distance-energy curves of graphite with AB-stacking as evaluated in single-point calculations using a hexagonal unit cell with 36 atoms and a fixed in-plane lattice constant of 2.47 Å (Figure S1). The minimization of the RMS error between FF and DFT energies yields $\varepsilon_{CA} = 2.906$ meV and $\sigma_{CA} = 3.340$ Å. The out-of-plane lattice constant at which the potential energy is minimal is 6.684 Å. The same LJ parameters are used for bulk sp$^2$-C atoms in graphitic structures (CG). As mentioned in Section II.2, CG–CG and CG–CA LJ interactions are purely repulsive, i.e., their LJ interaction is truncated at the energy minimum where the potential is shifted to zero.

### III.3 Bonded interactions between sp$^2$-C atom types

The parameters for bonded interactions between sp$^2$-C atoms are determined for CA atom types and then reused for the bulk atom types CB and CG. As a reference system, we use a graphene sheet in an orthorhombic cell with 128 atoms. We find a DFT-optimised lattice constant $a = 2.47$ Å, which directly yields $r_{eq,CA-CA} = \frac{a}{\sqrt{3}} = 1.426$ Å. The equilibrium angle $\theta_{eq,CA-CA-CA} = 120°$ is chosen according to graphene's hexagonal structure and the force constants for bond stretching $k_{r,CA-CA}$ and angle bending $k_{\theta,CA-CA-CA}$ are optimised using the elastic constants C$_{11}$, C$_{12}$ and C$_{66}$, and the objective function in Eq. ( 8 ) with $(i,j) \in \{(1,1),(1,2),(6,6)\}$ (Table 2). This optimisation step is carried out after step III.2 since the LJ interactions between CA atoms within the graphene sheet can affect its elastic constants, although this effect turns out to be negligible here.

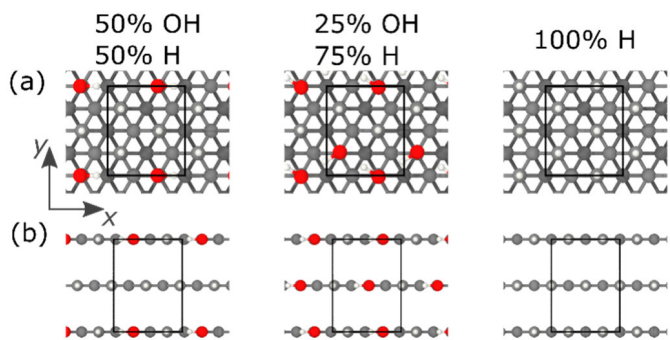

*Figure 3 Top views of the (a) diamond (111) surfaces and (b) graphite zigzag edges with 50% OH / 50% H, 25% OH / 75% H and 100% H terminations used for the parameter optimisation and testing. C, O, H atoms are shown as grey, red, and white spheres, respectively. Solid lines indicate the simulation cells.*

III.4 Charges of surface atom types

We use diamond (111) surfaces and graphite zigzag edges with different H and OH termination ratios as illustrated in Figure 3 as model systems for the optimisation of the parameters for surface atom types. In the following, we only use the percentage of hydrogen groups to refer to the different ratios, e.g., 75% H corresponds to a 25% OH / 75% H termination ratio. For the diamond (111) slabs, we use an orthorhombic cell with lengths 4.29 Å and 4.95 Å in the $x$ and $y$-directions, within which 48 C atoms are arranged in 6 atomic layers with four terminating species per surface. For the graphite edges, we use an orthogonal cell with lengths 4.92 Å and 6.64 Å in the $x$ and $y$-directions, 64 C atoms arranged in 8 atomic layers and four terminations per edge. All structures are initially relaxed such that the Euclidean norm of the DFT forces on each individual atom is lower than 0.01 eV/Å.

The charge parameters of C, O, and H surface atoms are optimised to fit the ESP above diamond surfaces and graphite edges with 75% H terminations. We use ESP charges[41,61–63] to ensure that, within a point-charge approximation, electrostatic interactions, that are represented by $E_{\text{Coulomb}}$, are separated by other nonbonded interactions, which are modelled by $E_{\text{LJ}}$. To fit the parameters, we use the REPEAT method[61] as implemented in CP2K[50]. Numerical details are given in Section S3 in the Supporting Information.

During the optimisation, we enforce local charge neutrality as in the original OPLS, i.e., the charge on each surface C atom is opposite to the total charge of its terminations. In order to ensure compatibility of the surface FF parameters with the other sets of OPLS parameters for liquids, we decided to adopt the charge parameters of H atoms in alkanes and benzene for the H terminations ($q_{H1} = 0.06e$ and $q_{H2} = 0.115e$)[34]. Tests show that this choice does not affect the results of the ESP fitting since the ESP near the 75% H-terminated surfaces is largely dominated by the charges of the OH groups. The charges of the atom types O1, H4, O2 and H5 are subsequently optimised yielding $q_{O1} = -0.59e$, $q_{H4} = 0.41e$, $q_{O2} = -0.44e$ and $q_{H5} = 0.42e$. A discussion on the accuracy of the fitted parameter set compared to the original OPLS charge parameter set for hydrocarbons and alcohols (Section S4) can be found in Section S5 in the Supporting Information.

Figure 4 shows examples of cross sections of $V_{DFT}$ and $V_{ESP}$ in the $x$-$y$-plane, 2.0 Å away from the surface atom with the highest $z$-coordinate for the diamond (Figure 4a) and the graphite case (Figure 4b). Additionally, the panels (c-f) illustrate one-dimensional $x$-$z$-cross-sections. It is interesting to observe that using the original OPLS charge parameters leads to the wrong periodicity of the ESP in the example shown in panel (d), while our optimised parameter yields improved results. This is caused by the larger magnitude of $q_{O1}$ in OPLS compared to our parameter set, which dominates the ESP in this specific region and conceals the contributions by nearby hydrogen and carbon atoms.

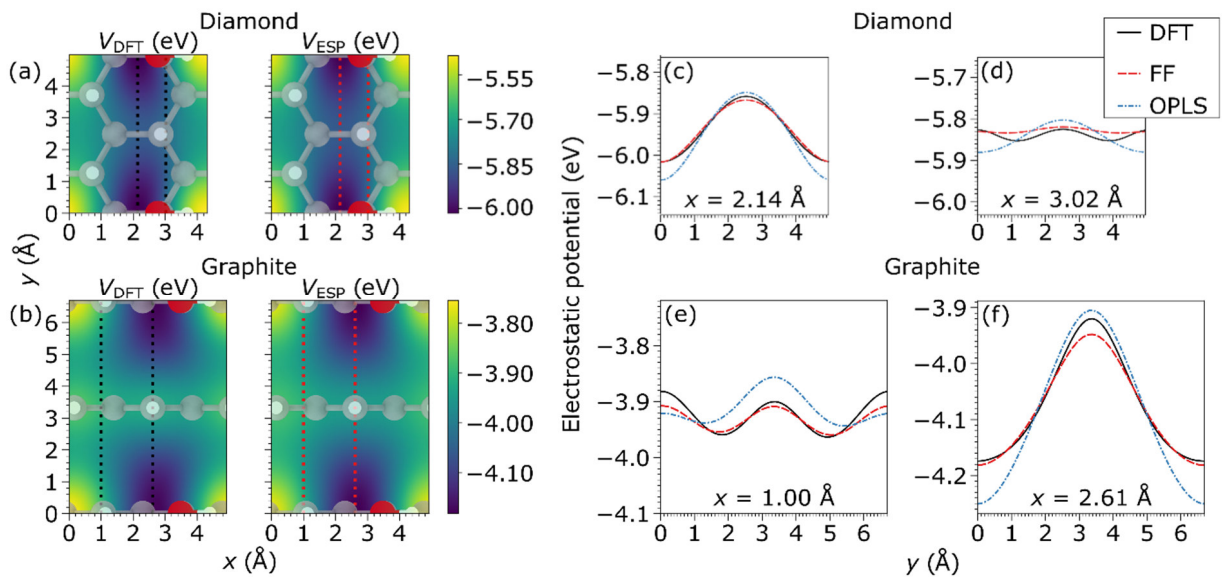

*Figure 4 Cross section of the ESP $V_{DFT}$ and $V_{ESP}$ calculated with DFT and our optimised charge parameter set on a diamond surface (a) and graphite edge that are passivated by 25% OH / 75% H. $V_{ESP}$ is shifted by an offset δ (Section S3) such that its average value coincides with the average of $V_{DFT}$. Panels (a) and (b) show a cross section through the x-y plane at a height of 2.0 Å above the atom with the largest z-coordinate. The black dotted lines in the $V_{DFT}$ plots and the red dotted lines in the $V_{ESP}$ plots indicate the one-dimensional cross sections shown in (c) and (d) for diamond, and in (e) and (f) for graphite. In (c-f), $V_{DFT}$ (black solid line) is shown in comparison to $V_{ESP}$ as obtained with optimised charges (red dashed line) and with the original OPLS charges (blue dash-dotted line).*

### III.5 Lennard-Jones parameters of surface atom types

As reference data for the optimisation of the LJ parameters of surface C, O and H atom types, we calculate energy-distance curves of diamond and graphite interfaces and of single $H_2O$ molecules adsorbed on these surfaces. This optimisation step is carried out after the optimisation of the charge parameters (Section III.4), so that the LJ potentials only model the nonbonded interactions that are not captured by the Coulombic interaction between point charges. To make the optimisation independent of the bonded parameters, we only vary the surface-surface and surface-molecule distances while keeping the $H_2O$ molecule and the slabs rigid. The structure of the rigid slabs and molecules corresponds to the minimum energy configurations of the surface-surface and surface-molecule interfaces as obtained by DFT geometry optimisation. Moreover, to make the atom types in the graphite model systems analogous to the atom types that will be used in the a-C target systems, we use CG rather than CA atom types for bulk graphite atoms. This choice does not affect the system geometry as the slabs are kept rigid.

In order to obtain reference structures for the surface-surface energy curves, we self-mate the systems of Figure 3 and create interfaces between each slab and a copy of it (Figure 5). For each surface type (Figure 3), we consider 2 to 4 different relative positions between the two slabs by translating the upper slab along the $x$ and $y$ directions (details in Section S6). Each system is then relaxed until the Euclidean norm of the DFT forces on each atom are lower than 0.01 eV/Å. In an analogous way, we create surface-water reference structures by placing and relaxing a single $H_2O$ molecule with different orientations and different lateral positions on the surfaces. The adhesion and adsorption curves are then calculated by translating the molecule or the upper slab rigidly along the $z$-direction from its equilibrium distance $z = z_{eq}$. All reported energy values are shifted so that the energy is zero at $z - z_{eq} = 20$ Å.

The generated data set consists of 11 surface-surface adhesion and 10 $H_2O$ adsorption curves for diamond, and of 8 adhesion and 9 $H_2O$ adsorption curves for graphite. For the optimisation of the LJ parameters for C1, H1, O1, CZ, H2 and O2, we select a subset of the energy-distance curves (Figure 5) and use the remaining data to test the transferability of the optimised parameters (Section IV). In line with the conventions used in the OPLS FF[34], we use the C1 LJ parameters for all remaining $sp^3$-C surface atom types and the CZ LJ parameters for the remaining surface $sp^2$-C atom types (Table 1), while we set $\varepsilon_{H4} = \varepsilon_{H5} = 0$. Numerical details about the optimisation and the definition of the objective function are provided in the Section S6.

Using a fit set consisting of 4 adhesion and 3 $H_2O$ adsorption curves for diamond (i.e., for the atom types C1, H1 and O1), and 4 adhesion and 2 $H_2O$ adsorption curves for graphite (i.e., for the atom types CZ, H2 and O2) we obtain several different parameter sets that can accurately reproduce the reference data. Out of these sets, we choose the set of parameters that are closest to the original OPLS values for the respective groups in hydrocarbons, alcohols and phenol[34] in order to achieve good compatibility between the present FF and OPLS and thus facilitate future extensions of the FF to other lubricants.

Figure 5 shows the complete fit set of adhesion and $H_2O$ adsorption configurations. For each configuration, the DFT energy curve is plotted as a function of the displacement $z - z_{eq}$ from the relaxed configuration. In addition, we show the curves as obtained by calculations with the optimised FF parameters, along with results calculated with the original OPLS LJ parameters and charges (see Section S4 for a full list of the parameters). The parameter optimisation leads to an overall improved description of most diamond adhesion configurations (Figure 5a) compared to the original OPLS parameters. The only exception is the first configuration for 50% H surface termination, where the potential depth is slightly overestimated by 5.6 meV/Å$^2$ by the FF with optimised parameters while the original OPLS parameters yield an underestimation of 3.6 meV/Å$^2$. In the other cases the maximal absolute deviation from the DFT energy minimum is 2.8 meV/Å$^2$ for the optimised FF and 8.9 meV/Å$^2$ for the OPLS parameters. For graphite interfaces (Figure 5b), both the FF with optimised parameters and the original OPLS parameters can accurately describe the adhesion energies with a largest minimum energy deviation of 1.2 meV/Å$^2$ for the FF (50% H) and of 2.3 meV/Å$^2$ for the OPLS parameters (100% H). The FF description of $H_2O$ adsorption on diamond (Figure 5c) and graphite (Figure 5d) also shows slight improvements with respect to the original OPLS parameters. However, to further improve the surface-water interactions, a more complex model than the simple TIP3P water model would be required (see Section VI).

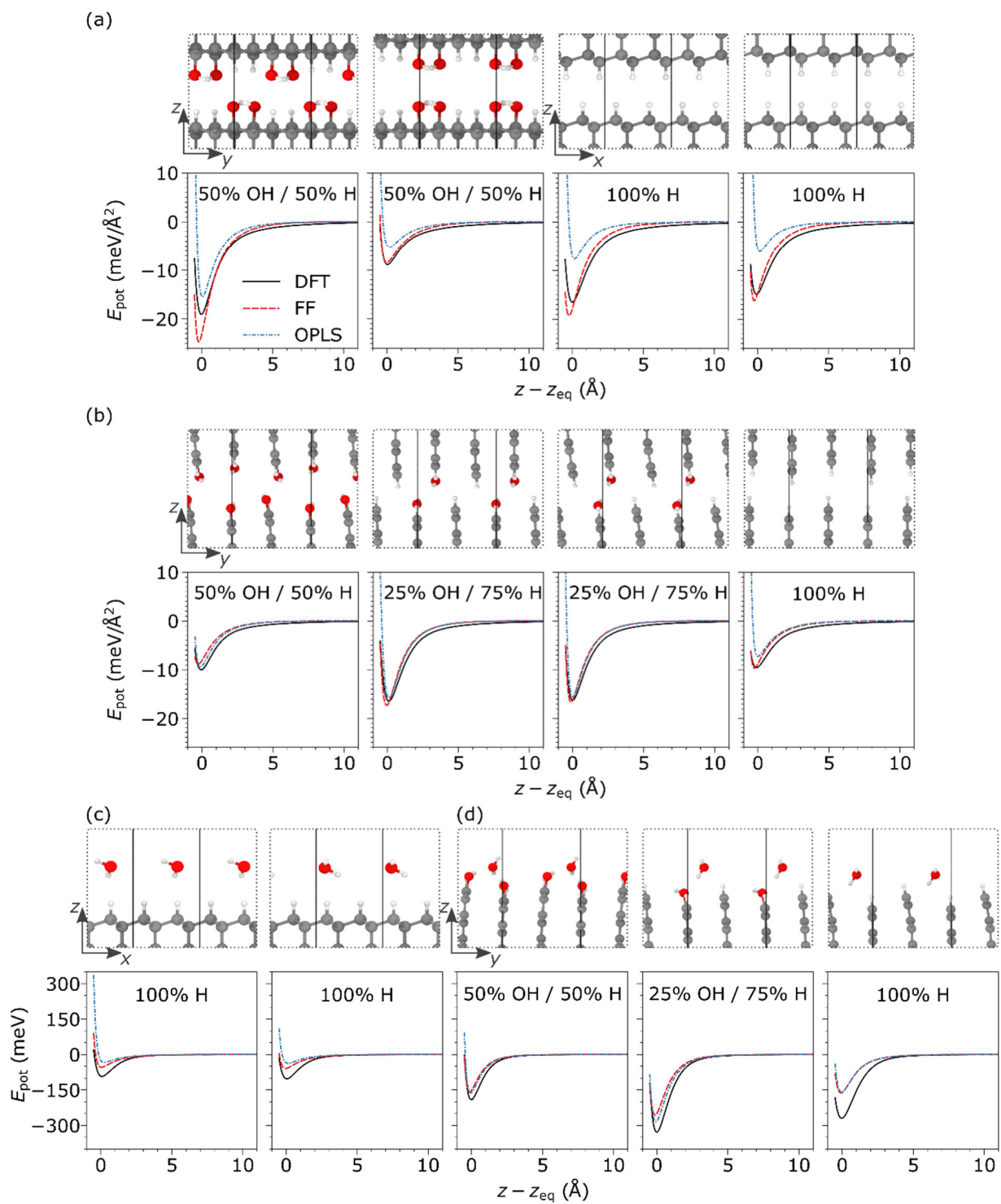

*Figure 5 Set of model systems and data used for the optimisation of the LJ parameters of surface C, O, H atom types. The data consists of rigid adhesion curves of (a) self-mated diamond surfaces, (b) self-mated graphite edges, and of (c) diamond-$H_2O$ and (d) graphite-$H_2O$ adsorption curves. Adhesion energies are given per interface area and $H_2O$ adsorption energies are given per $H_2O$ molecule. The three curves shown in each diagram indicate the results of the reference DFT calculations (black solid lines) and of FF calculations with optimised parameters (red dashed lines) and original OPLS parameters (blue dash-dotted lines). Pictures of the corresponding structures in the respective DFT energy minimum are shown above each diagram. Solid lines in the atomic structure pictures show the simulation cell. See Supporting*

*Information for the full set of fit and test configurations and details about the relative lateral positions of the surface-surface and surface-water systems.*

### III.6 Bonded interactions between surface atom types

In the last step, we optimise the parameters of the bonded interactions between surface terminations. As reference data, we use DFT energy variations when stretching individual surface bonds and bending individual surface angles while keeping the remaining atoms fixed to their relaxed DFT position. Bond stretching is performed along the direction of the vector between the two atoms to keep all other bonded interactions constant. During angle bending the lengths of the bonds that enclose the angle are kept fixed. Dihedral energy variations during angle bending turn out to be negligible for our fit data. The parameters obtained in this optimisation step depend on the results of the optimisation steps III.2, III.4 and III.5 since the atoms displaced upon bond stretching/angle bending interact with other atoms in the system via LJ and Coulomb potential terms.

To optimise the parameters $r_{eq}$, $k_r$ and $\theta_{eq}$, $k_\theta$ for all $sp^3$-C–H1 bonds and C–$sp^3$-C–H1 angles described in Figure 1b, we perform C1–H1 bond stretching (example in Figure 6a) and CD–C1–H1 angle bending on the fully H-passivated (111) diamond surface (Figure 3). The $sp^3$-C–O1 and O1–H4 bond parameters, and the C–$sp^3$-C–O1 and $sp^3$-C–O1–H4 angle parameters are optimised using the diamond (111) surface with 75% H passivation. We use the fully H-terminated graphite zigzag edge for $sp^2$-C–H2 bond and C–$sp^2$-C–H2 angle parameters. CY–O2 and O2–H5 bond parameters, and CY–O2–H5 (example in Figure 7a) and C–CY–O2 angle parameters are optimised using the 75% H-terminated graphite edge.

The FF parameters in this step are optimised using a least-squares fit. The parameter $\theta_{eq}$ for all C–O–H angles is treated as a fit parameter, while the equilibrium value for the other angles is set to 120° for all angles centred on $sp^2$-C atoms, and to 109.5° for all angles centred on $sp^3$-C atoms in line with the conventions used in the original OPLS FF[34]. This choice leads to a slight overestimation of the minimum energy angle for CD–C2–O1 and CG–CY–O2 angles. Therefore, in these two cases we only fit the curvature of the energy curve. Figure 6b and Figure 7b show the energy curves obtained with DFT and optimised FF for bond stretching and angle bending, respectively. Energy curves calculated with the original OPLS bond and angle parameters for hydrocarbons, alcohols[55] and phenol[64] are also reported for comparison (blue curves).

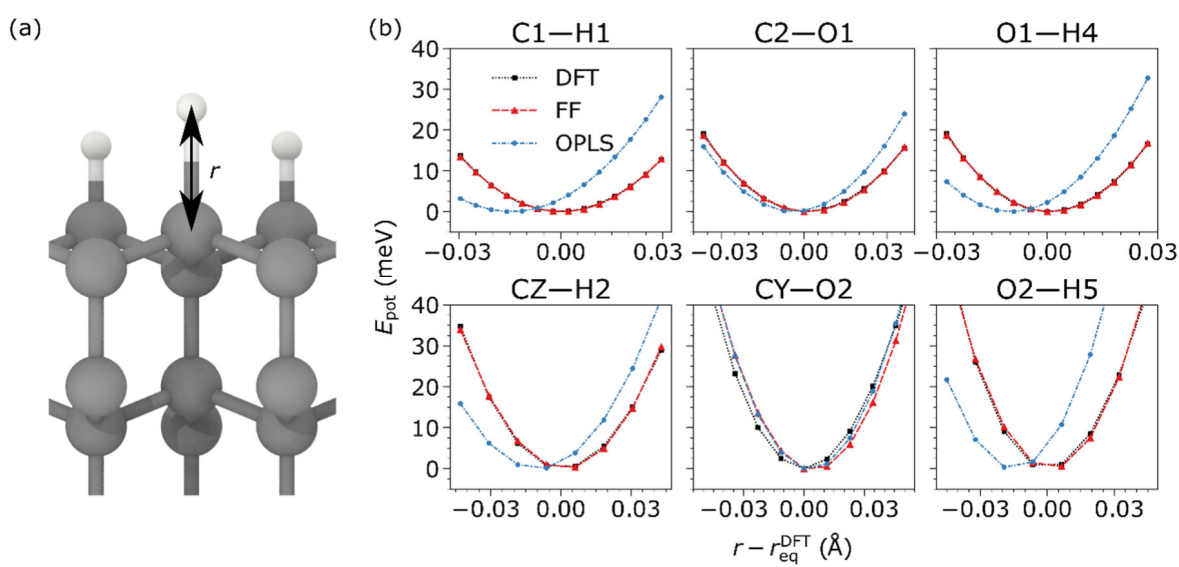

*Figure 6 Optimisation of surface bond-stretching parameters. (a) Stretched C1–H1 bond on a diamond (111) surface with 100% H terminations. (b) Energy as a function of bond length for different surface bonds computed with DFT (black squares), with the optimised FF (red triangles) and with the original OPLS parameters (blue circles). The energies are plotted as a function of $r-r_{eq}^{DFT}$, where $r$ is the bond length between the atom types indicated on top of each diagram and $r_{eq}^{DFT}$ is the corresponding bond length in the DFT energy minimum.*

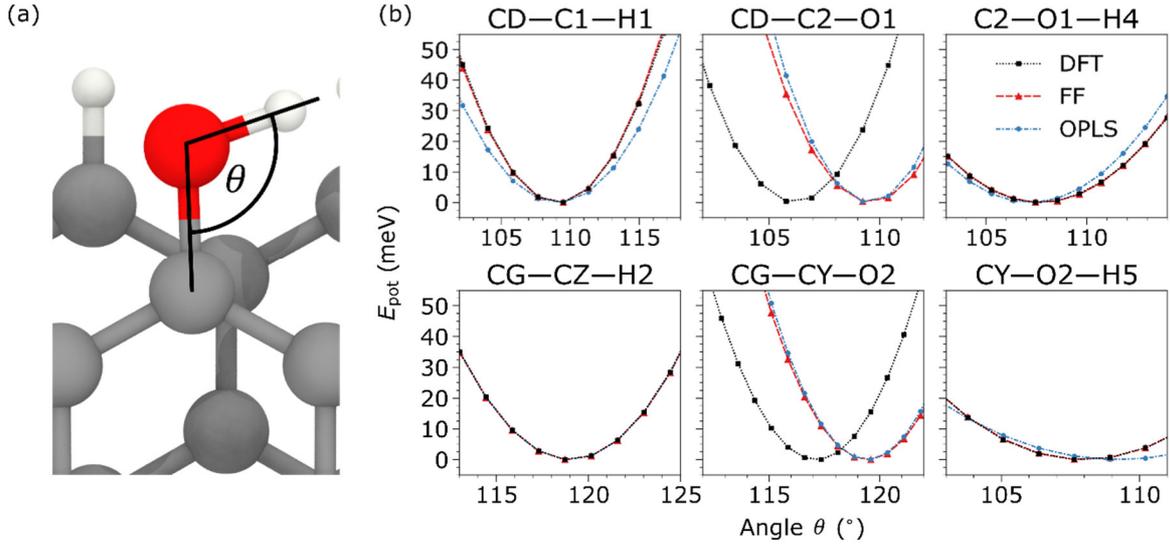

*Figure 7 Optimisation of surface angle-bending parameters. (a) CY–O2–H5 angle on a 25% OH / 75% H-terminated graphite edge. (b) Energy as a function of the bond angle for different surface angles computed with DFT (black squares), with the optimised FF (red triangles) and with the original OPLS parameters (blue circles). The energies are plotted as a function of the angle enclosed by the atom types indicated on top of each diagram.*

## IV. Testing

To test the transferability of the optimised parameter set, we consider three sets of test systems: crystalline surfaces that were not part of the fitting set (Section IV.1, S6); bulk a-C (Section IV.2); a-C surfaces (Section IV.3). For the H- and OH-passivated crystalline surfaces we test the ESP (Section S6), interface adhesion and water adsorption (Section IV.1). In Section IV.2, we test the transferability of the parameter set to bulk a-C systems by calculating their mechanical properties for different a-C densities. Finally, in Section IV.3, we test the accuracy of the parameter set in describing dry adhesion and water adsorption on a-C surfaces. Here, we also present tests on the potential energy corrugation of a-C/a-C and a-C-$H_2O$ interfaces, because of the relevance of this quantity in friction. In all test cases, we consider system sizes that allow us to evaluate these quantities by means of DFT and the FF on the same model systems.

IV.1 Crystalline surfaces

**Interface adhesion and H₂O adsorption with rigid systems.** Here we use model systems that were not used for Step (5) of the parameter optimisation (Section III.5). A full overview and more details of these test systems are provided in the Supporting Figs. S2-S5. Figure 8 shows a selection of interface adhesion and water adsorption curves obtained using rigid systems for graphite zigzag edges and diamond (111) surfaces with 75% H terminations. We observe that the FF with optimised parameters can accurately reproduce the DFT test data and, in general, improve upon the original OPLS parameters. In Figure 8a-d the absolute deviations from the DFT energy minimum are 0.7 meV/Å$^2$, 0.3 meV/Å$^2$, 103 meV, and 94 meV for the FF and are 8.4 meV/Å$^2$, 1.7 meV/Å$^2$, 162 meV, and 86 meV for the OPLS parameters, respectively. The very limited improvement in the water adsorption curves can be ascribed to the limitations of the TIP3P model and, in particular, to the absence of LJ interactions for the H atoms in the water molecules. This aspect will be discussed in more detail in Section VI.

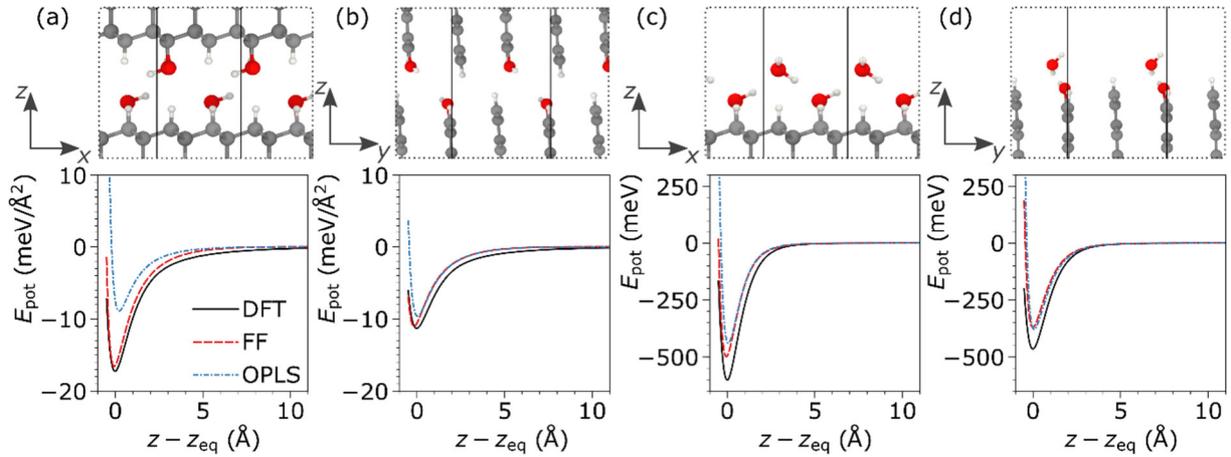

*Figure 8 Test cases of interface adhesion and H₂O adsorption with rigid systems (analogous to to Figure 5): (a) diamond/diamond adhesion, (b) graphite/graphite adhesion, (c) H₂O-diamond adsorption, (d) H₂O-graphite adsorption. All surfaces and edges are terminated with 25% OH / 75% H. The full set of test configurations and information about the lateral positions are provided in the Supporting Information.*

**Interface adhesion and H$_2$O adsorption energies with relaxed geometries.** As a stricter test, we calculate interface adhesion and H$_2$O adsorption energies, where we lift the constraint of rigid structures. To determine diamond/diamond and graphite/graphite adhesion energies, we self-mate the systems from Figure 3 and consider four different lateral positions for each system. For these calculations, the simulation cells of Figure 3 are replicated three times along the $x$ and $y$ directions in the case of diamond and two times in the case of graphite. We relax these systems until all force components on all atoms are lower than 0.01 eV/Å (see Section S7 for snapshots of the relaxed DFT structures and for details about geometric constraints applied to the outermost atoms of the slabs during the geometry optimisation). Afterwards, each slab is relaxed individually. The adhesion energy is calculated as the difference between the energy of the relaxed, self-mated system and the sum of the energies of the relaxed slabs. To calculate the water adsorption energy on these surfaces/edges, we place a single, randomly oriented H$_2$O molecule on top of the surface/edge in four different lateral positions. The adsorption energy is calculated analogously to the previously described adhesion energies.

In Figure 9, we compare the interface adhesion and H$_2$O adsorption energy results as computed with DFT (black), with the optimised FF (red) and with the FF with original OPLS LJ parameters and charges (blue). For all considered diamond surface terminations, the optimised parameter set is more accurate than the original OPLS parameter set. More importantly, the ranking of the surface adhesion energies (Figure 9a) for different surface terminations is the same in DFT and the FF, while the original OPLS parameters underestimate the adhesion between 100% H-terminated diamond surfaces. As we already mentioned when discussing the rigid adsorption curves, the FF and OPLS description of the H$_2$O adsorption on diamond (Figure 9b) is limited by the functional form of the employed water model (see Section VI). Nonetheless, the DFT ranking of water adsorption energies is preserved by both FFs. For polar surfaces with OH terminations the optimised FF is rather accurate and closer to DFT than the original OPLS FF. Both FFs underestimate the adsorption energies on the hydrophobic 100% H-terminated diamond surface. Similar considerations apply to the interface adhesion (Figure 9c) and water adsorption energies (Figure 9d) on graphite edges. In these cases, however, the energies calculated with the original OPLS nonbonded parameters for hydrocarbons and alcohols are already accurate for edges with OH polar groups and an improvement by the optimised FF can only be observed in the case of fully hydrogenated edges. Here, the absolute deviation from DFT reduces from 2.96 meV/Å$^2$ to 0.8 meV/Å$^2$ for the adhesion and from 130 meV to 101 meV for the H$_2$O adsorption.

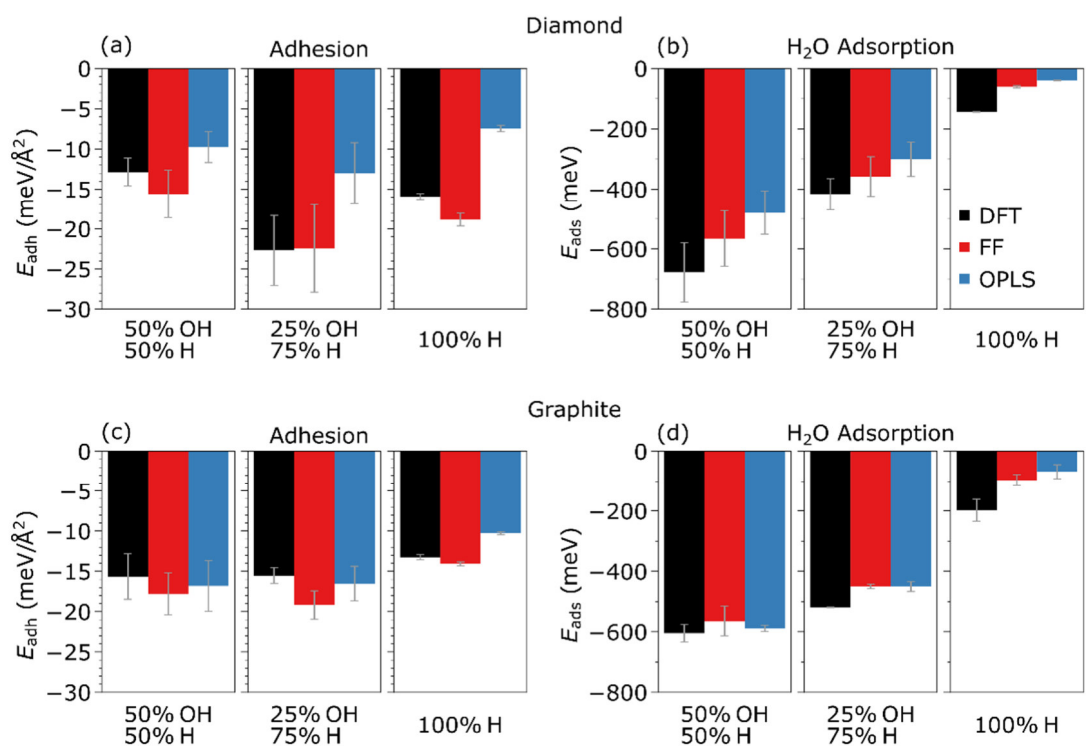

*Figure 9 (a) Diamond-diamond adhesion energy, (b) H$_2$O adsorption energy on diamond, (c) graphite-graphite adhesion energy, and (d) H$_2$O adsorption energy on graphite. The bar diagrams show the arithmetic mean of the energies calculated at four different lateral positions of the upper structure (H$_2$O or slab). Error bars show the standard deviation of the mean. The plots show the values calculated with DFT (black bars), with the FF using optimised parameters (red bars) and with the FF using original OPLS surface LJ parameters and charges (blue bars). All adhesion energies $E_{adh}$ are given per interface area and H$_2$O adsorption energies $E_{ads}$ are given per H$_2$O molecule. The surface terminations are indicated below each diagram.*

## IV.2 Bulk amorphous carbon

In our modelling strategy for bulk amorphous carbon (Section II.2), the FF parameterisation is based solely on data obtained from crystalline solids. To test how successful this strategy is in describing the structure and elastic properties of a-C, we generate a set of 32 bulk a-C structures using quench-from-the-melt simulations and the screened Tersoff potential[65,66] (details of the protocol are described in the Supporting Information S8). The structures have varying densities ranging from 1.75 g/cm$^3$ to 3.5 g/cm$^3$. We calculate their bulk modulus, Young's modulus and Poisson's ratio from the slope of stress-strain curves in analogy to the calculations of Table 2. Figure 10 shows the three elastic moduli computed with DFT and with the optimised FF model as a function of the a-C density. The FF calculations accurately match the results from DFT calculations over the entire range of considered densities. This is remarkable given the simple FF description of bulk a-C using only three different atom types CD, CG and CB with parameters fitted to crystalline phases. The FF performs particularly well for high a-C densities, where the structures are dominated by sp$^3$-C atoms. For low densities, the FF calculations of the bulk modulus (Figure 10a) and Young's modulus (Figure 10b) agree well with the DFT values, while the FF slightly underestimates the Poisson's ratio (Figure 10c).

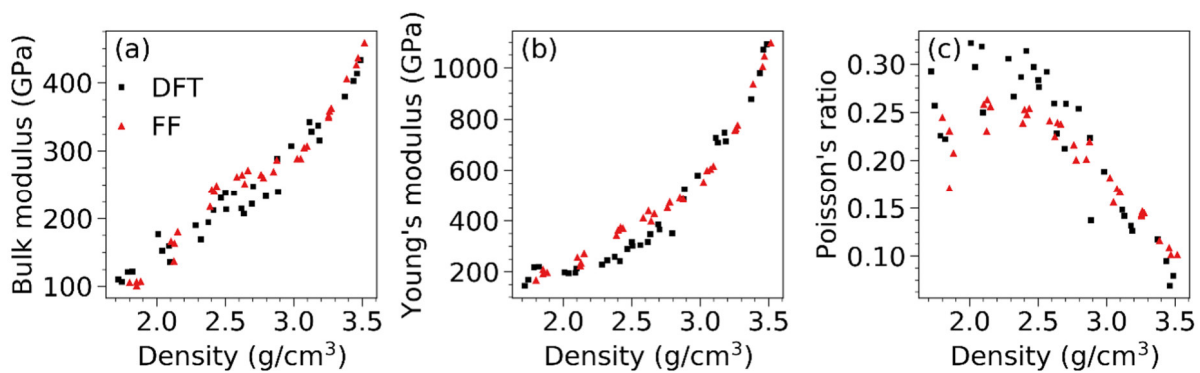

*Figure 10 Elastic moduli of bulk a-C samples as calculated with DFT (black squares) and with the FF with optimised parameters (red triangles). Panel (a) shows the bulk modulus, panel (b) the Young's modulus and panel (c) the Poisson's ratio as a function of the a-C density.*

## IV.3 Amorphous carbon surfaces

Interface adhesion and H$_2$O adsorption energies with relaxed geometries. To test interface adhesion and H$_2$O adsorption energies for a-C surfaces, we create three slab models using bulk systems with densities 2.0 g/cm$^3$ 2.5 g/cm$^3$ and 3.0 g/cm$^3$. For each slab model we consider three different surface terminations: 100% H, 75% H, and 50% H. After a relaxation of the unpassivated slab surfaces, H terminations are introduced by iteratively placing individual H atoms in the vicinity of surface dangling bonds and relaxing the system using the density-functional tight-binding method[67] to calculate interatomic interactions. Only if the adsorption is energetically favourable with respect to a H$_2$ reservoir, the termination is accepted. OH-terminated systems are generated by replacing randomly selected H terminations of the 100% H-terminated case with OH. Details can be found in the Supporting Information S8. The calculation of the adsorption and adhesion energies of each surface/surface and surface/water system follows an analogous protocol as previously discussed for the diamond case in Section IV.1 (see Section S9 for snapshots of the DFT relaxed structures).

The results of these calculations are presented in Figure 11. For all three densities, both the ranking of the adhesion and of the H$_2$O adsorption energies is correctly predicted by the optimised FF (within the typical scattering of different relative positions as indicated by the error bars). In contrast, the original OPLS parameters clearly give an underestimation of most adhesion and adsorption energies. The only exception is the H$_2$O adsorption on the a-C surface with 50% H terminations and density 3.0 g/cm$^3$ (Figure 11c). Here, the FF with optimised parameters overestimates the adsorption energy by about 0.1 eV compared to both DFT and the original OPLS parameters. Regarding the adsorption of a H$_2$O molecule on fully H-terminated a-C surfaces, we observe the same behaviour as in the previous sections: the FFs underestimate the H$_2$O adsorption energies, irrespective of the parameter set used.

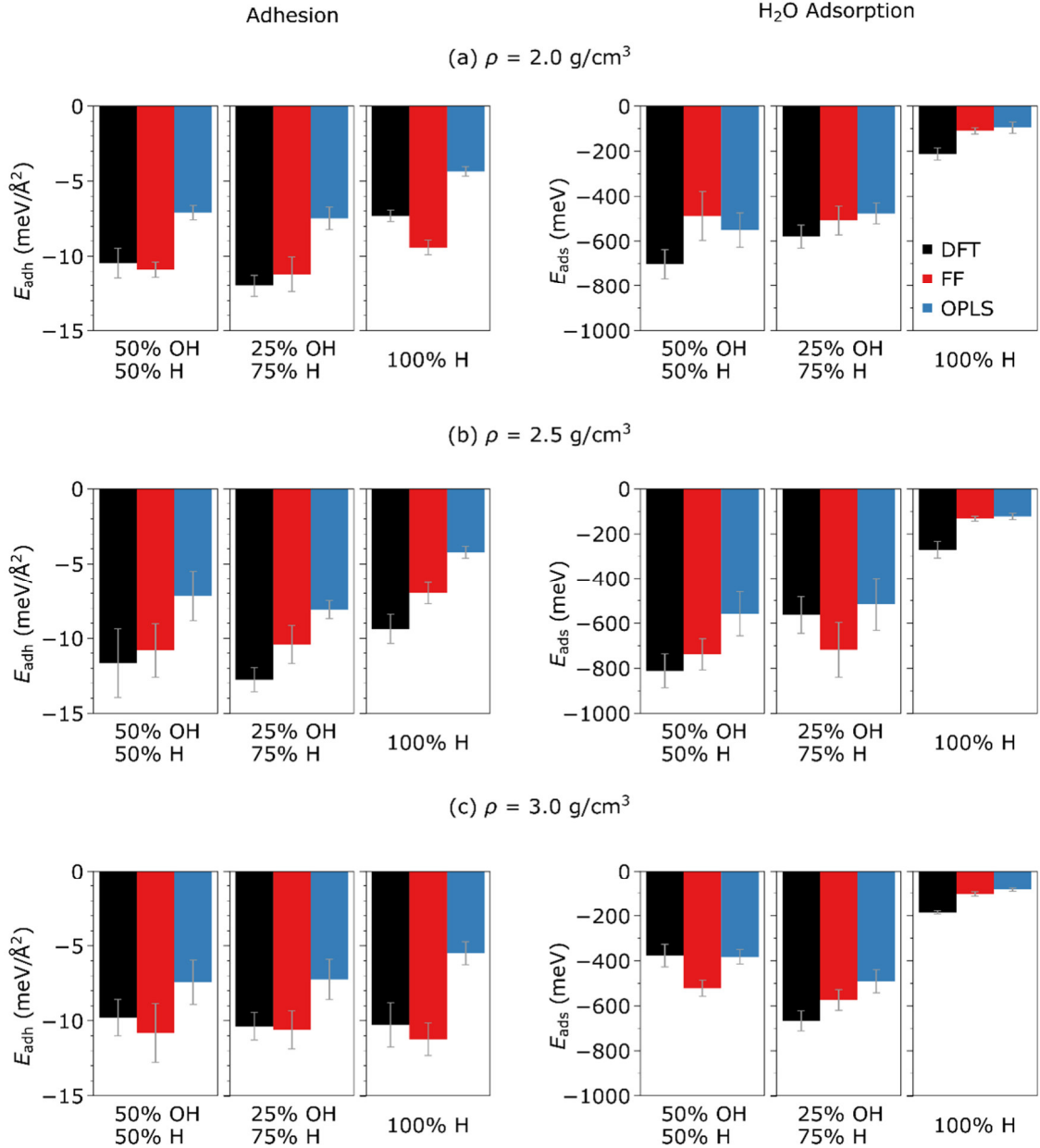

*Figure 11 a-C/a-C adhesion (left column) and a-C/$H_2O$ adsorption (right column) energies. The three rows show results for three a-C densities: (a) 2.0 g/cm³, (b) 2.5 g/cm³, (c) 3.0 g/cm³. The bar diagrams show the arithmetic mean of the energies calculated at four different lateral positions of the upper structure ($H_2O$ or slab) using DFT (black), the FF with optimised parameters (red) and with the original OPLS LJ parameters and charges (blue). Error bars show the standard deviation of the mean. Adhesion energies $E_{adh}$ are given per interface area and $H_2O$ adsorption energies $E_{ads}$ per $H_2O$ molecule.*

**Potential energy corrugation of a-C interfaces.** The main goal of the development of this FF is the simulation of tribological a-C interfaces. This requires an accurate description of the elastic properties of a-C, as demonstrated above, as well as of the potential energy landscape generated by the relative motion of the two interacting surfaces, i.e., the energy barriers that must be overcome during sliding[24].

For the dry case, we consider four different interfaces by pairing two a-C slabs with the same density (2.0 g/cm$^3$ or 3.0 g/cm$^3$) and the same termination ratio (75% H or 100% H) as shown in Figure 12a-d. We reuse the systems presented in Figure 11, and start from the DFT-relaxed interface with the most energetically favourable relative position of the slabs in each case. We calculate one-dimensional potential energy profiles by moving the upper surface relative to the lower surface in increments of 0.12 Å along the $x$ direction (Figure 12) in the absence of normal load. During this procedure the individual slabs are kept rigid to avoid elastic instabilities (which would prevent a direct comparison between DFT and FF results). After each displacement, the $z$ position of the upper slab is optimised.

Figure 12 shows the results of these simulations as obtained by DFT and by the FF with optimised parameters and with original OPLS parameters for hydrocarbons and alcohols[34]. Panels (e-h) show the variation of the potential energy as a function of the displacement in $x$-direction relative to the potential energy at the initial position. Panels (i-l) show the corresponding variation of the optimised upper slab position $z$ relative to the initial position $z_{eq}$ ($\Delta z = z - z_{eq}$), i.e., the geometric interface corrugation.

The potential energy landscape of both FF calculations agrees with the DFT results within a maximum absolute error of 3.2 meV/Å$^2$ for the optimised parameter set and of 2.5 meV/Å$^2$ for the OPLS parameter set. All three energy profiles show the same number of energy minima and maxima at about the same relative positions. For all four test systems, the geometric corrugation (Figure 12i-l) obtained using the optimised parameter set is closer to the DFT results than the corrugation obtained with the original OPLS parameters (maximum deviations of 0.4 Å and 0.7 Å, respectively).

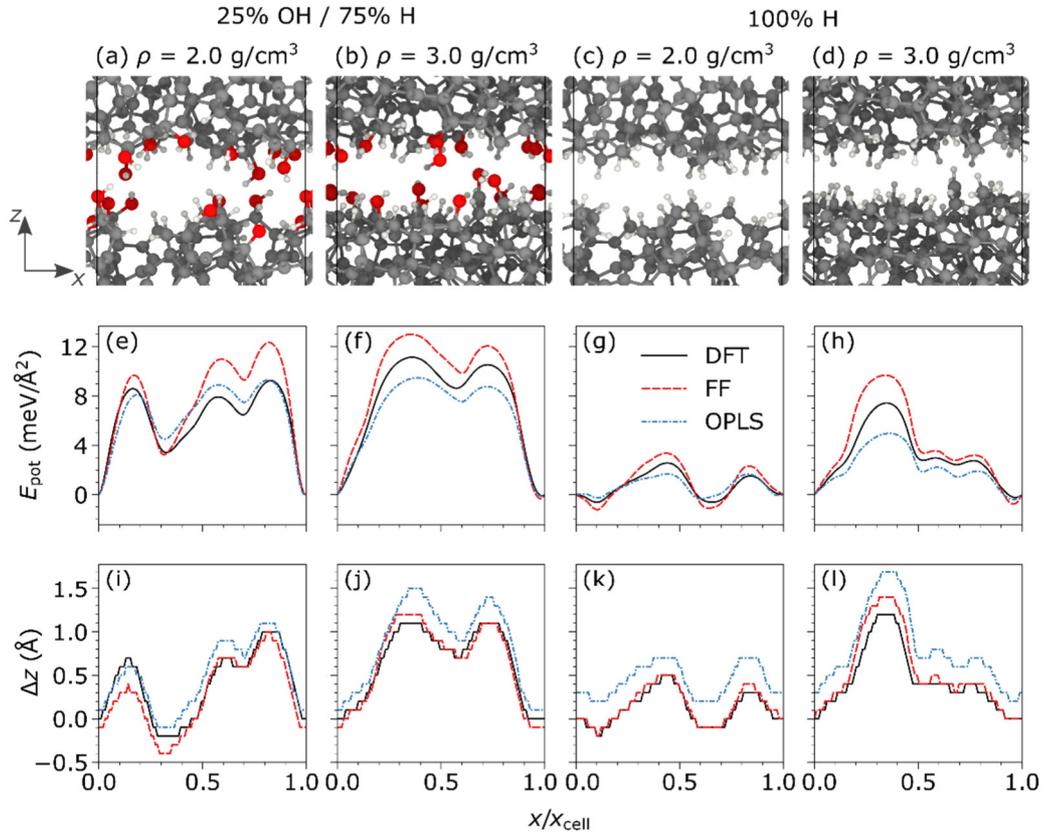

*Figure 12 Energy and geometric corrugation profiles for dry interfaces between rigid a-C systems. (a-d) Snapshots of the interfaces at the initial relative position ($x = 0$) after DFT geometry optimisation. (e-h) Variation of the minimised energy obtained by moving the upper slab quasistatically along the x direction for one unit cell length $x_{cell}$: DFT (black solid line), optimised FF (red dashed line) and original OPLS FF (blue dash-dotted line). After each displacement step, the z position of the upper rigid slab is optimised. The panels (i-l) show the corresponding height profiles relative to the initial optimised z position $z_{eq}$: $\Delta z = z - z_{eq}$.*

Devising analogous tests on interfaces that are in boundary lubrication with water (i.e., from less than a water monolayer to a few water monolayers) is not trivial. Here we simplify the problem and perform a well-defined test. We calculate one-dimensional energy profiles of the interaction between one $H_2O$ molecule and the a-C slabs (Figure 13a-d). The simulations follow an analogous protocol as for the dry a-C/a-C corrugation calculations. We start from the most favourable adsorption position of the $H_2O$ molecule on the surfaces as obtained with DFT (Figure 13a-d), we keep both the H$_2$O and the slab rigid, and calculate the optimal height of the H$_2$O molecule and the corresponding energy for each step along the $x$ "sliding" path.

We find that the two sets of FF parameters yield very similar results. The potential energy (Figure 13e-f) tends to be slightly underestimated by the FFs, while the position of the minima and maxima along the $x$ direction is reproduced fairly well. The maximum energy deviation from DFT is 0.30 eV for the optimised FF and 0.25 eV for OPLS. The geometric corrugation (Figure 13i-l) is well-described by the FFs with a maximal deviation of 0.5 Å for the optimised FF and of 0.4 Å for the OPLS parameters.

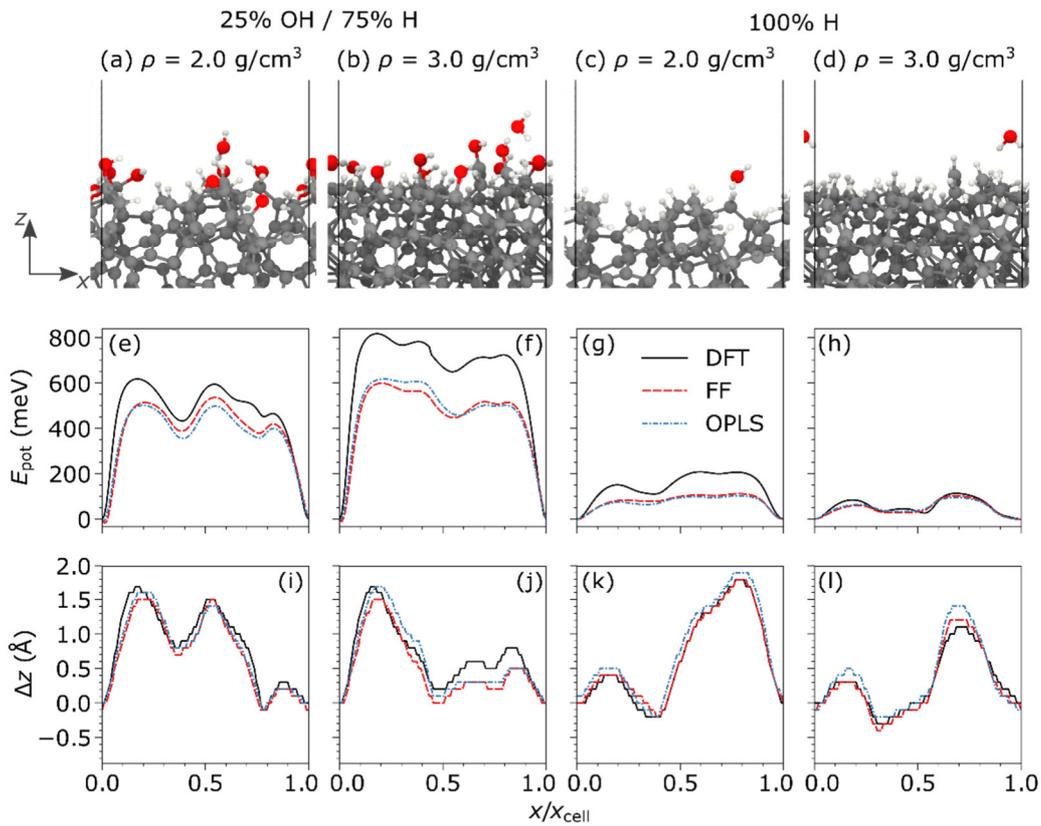

*Figure 13 Energy and geometric corrugation profiles for a rigid H$_2$O molecule moving along the x direction on different a-C surfaces. (a-d) Snapshots of structures at x=0 and $z = z_{eq}$ after geometry optimisation with DFT. $z_{eq}$ is the initially optimised z position of the H$_2$O molecule.*

*(e-h) Energy variations for the molecule that moves quasistatically along the x direction by one unit cell length $x_{cell}$: DFT results (black solid line); results obtained with the optimised FF (red-dashed lines) and with the original OPLS FF (blue dash-dotted line). After each displacement step, the z position of the rigid $H_2O$ molecule is optimised. Panels (i-l) show the corresponding height profile relative to $z_{eq}$.*

## V. Example applications

Despite the relative simplicity of the FF and the fact that its parametrization is based on well-defined crystalline reference systems, testing the optimised FF against DFT reveals its transferability out of the fitting set. In most cases, the optimised FF (and often also the original OPLS FF) show quantitative agreement with DFT calculations of elastic, adhesion and adsorption properties of a-C/a-C and a-C/water interfaces. In this section, we increase the system sizes and simulated times, thereby renouncing a direct comparison with DFT, and present two examples of how the FF can be used to gain insights into tribological processes. First, in Section V.1, we perform dynamic sliding simulations of dry a-C/a-C interfaces and look for relationships between the chemical structure of the passivated interface and the shear stress. Next, in Section V.2, we consider water-lubricated a-C/a-C interfaces and investigate how the shear stress varies as a function of the water film thickness.

### V.1 Relationship between dry friction and corrugation of the contact potential energy surface

The relatively low computational cost of the FF-based simulations enables an explicit calculation of the shear stress in sliding MD simulations. As a result, a direct investigation of the relationship between friction and the chemical structure of the tribological interface becomes possible in model simulations. A correlation between friction and the corrugation of the contact potential energy surface (CPES; the potential energy landscape due to the surface-surface interaction[24]) is predicted by theory[23] and has been recently verified using a combination of MD and quasistatic simulations on H- and F-passivated diamond and a-C surfaces[24]. The present development allows us to further investigate this correlation for a-C systems with partial OH passivation, which is often reported to result from the dissociative adsorption of water or other OH-containing molecules[4]. This is particularly interesting since surface OH groups add complexity to the problem owing to their additional bending and rotational degrees of freedom that can contribute to energy dissipation.

For this example application, we create interfaces between two a-C slabs with a density $\rho = 2.5$ g/cm$^3$ (Figure 14a) and consider three different surface termination types 50% H, 75% H and 100% H (Figure 14b). The preparation of the systems closely follows the protocol described in Section IV.3, but we use the screened REBO2 potential[65,68] to introduce H terminations into the system (see Supporting Information S8 for the details). Each slab has a height of about 50 Å and the interface area is 20 Å × 20 Å, resulting in a total system size between 5240 and 5378 atoms. We use a timestep of 0.5 fs to integrate the equations of motion. A normal load of $P_N = 1$ GPa is imposed using the pressure-coupling algorithm developed by Pastewka et al.[18] and the temperature is controlled using a Langevin thermostat with a damping constant of 0.1 ps far from the interface as explained in detail in Ref. 18. After an initial equilibration of the normal pressure and of the temperature (1 GPa and 300 K, respectively), we switch off the thermostat[24] and apply a constant sliding speed $v = 10$ m/s to a rigid layer of atoms with a height of 6 Å on the top of the upper slab. Another layer of atoms with the same height on the bottom of the lower slab is kept fixed at its initial position. Each sliding simulation runs for 2 ns. After a running-in period of 0.5 ns, we calculate the average shear stress in three time intervals from 0.5 ns to 0.9 ns, 1.05 to 1.45 ns, and 1.6 to 2 ns. We denote the average of these three measurements with $\langle\tau\rangle_\text{MD}$, and estimate its error with the standard deviation of the mean.

To relate $\langle\tau\rangle_\text{MD}$ to the corrugation of the CPES, we follow Ref. 24 and use an estimate of the sum of the energy barriers experienced by the systems during sliding as a measure of the corrugation. Under the assumption that the energy that is required to overcome the barriers is completely dissipated, this sum normalised by the sliding distance corresponds to an idealised shear stress, which we call $\langle\tau\rangle_{\text{ideal}}$ in the following. For this purpose, we calculate the potential energy profile along the respective sliding direction by quasistatic shear simulations, in which the upper fixed layer is moved in increments of 0.1 Å along this direction. After each step, the system is relaxed to a local potential energy minimum using a threshold of $10^{-4}$ eV/Å for the Euclidean norm of the force vector on all atoms. During this relaxation the upper and lower rigid layers remain fixed. The height of the system is kept constant at the corresponding average height measured in the MD sliding simulation during the time interval from 0.5 ns to 2 ns. The maximal deviation from this average height during the dynamic simulation in this time interval is smaller than 0.04 Å for all systems. To avoid the bias caused by the initial arbitrary relative configuration of the slabs, we first slide for an initial "running-in" period corresponding to a sliding length $L = 20$ Å (i.e., the length of the simulation cell in $x$- and $y$-direction). Afterwards, we calculate the potential energy profile for three additional cycles over the simulation cell (i.e., an additional sliding distance $3L$). For each of these cycles, we calculate the sum of the energy barriers and normalise it by dividing the energy sum by $L$. We denote the arithmetic mean of the three values with $\langle\tau\rangle_{\text{ideal}}$, and use the standard deviation of the mean as an error estimate.

For each of the three termination types, we consider two different a-C samples with the same density and two different sliding directions (the $x$- and $y$-directions of the simulation cell). The results are presented in Figure 14c and show a close correlation between the average shear stress $\langle\tau\rangle_{\text{MD}}$ from MD simulations and $\langle\tau\rangle_{\text{ideal}}$ from quasistatic simulations. There is a clear ranking between the three surface termination ratios: the shear stress increases with increasing OH termination ratio. This is presumably caused by the formation of hydrogen bonds across the interface in combination with an increased steric hindrance between the OH terminations of the two surfaces. Similar findings are presented in Ref. 21 for a-C surfaces with chemisorbed linear alkanes and alcohols. Moreover, also the scattering and the error bars of $\langle\tau\rangle_{\text{ideal}}$ and $\langle\tau\rangle_{\text{MD}}$ tend to increase with increasing OH content, probably because of the bending and rotational degrees of freedom of the OH groups.

These results also prompt three more general considerations. Firstly, the correlation between shear stress and CPES corrugation confirms that the sum of the energy barriers experienced during sliding is a central quantity that determines friction for these systems. This suggests that testing the accuracy of a FF for tribological applications in reproducing the CPES, as in Figure 12, is crucial for the reliability of tribological simulations. Secondly, the $\langle\tau\rangle_{ideal}$ values are generally larger than the corresponding $\langle\tau\rangle_{MD}$ values indicating that not all the energy that is required to overcome energy barriers is dissipated. Future studies using this FF could target this aspect to elucidate the mechanisms underlying energy dissipation in dry carbon contacts[69]. Finally, the ranking and magnitude of the shear stresses in Figure 14 indicates that a 100% H surface termination is more suitable than a mixed OH/H termination to achieve superlubricity in dry a-C contacts. This suggests that in presence of oxygen, superlubric interfaces are likely terminated by groups other than OH, for instance by ether, hydrogen or aromatic structures.

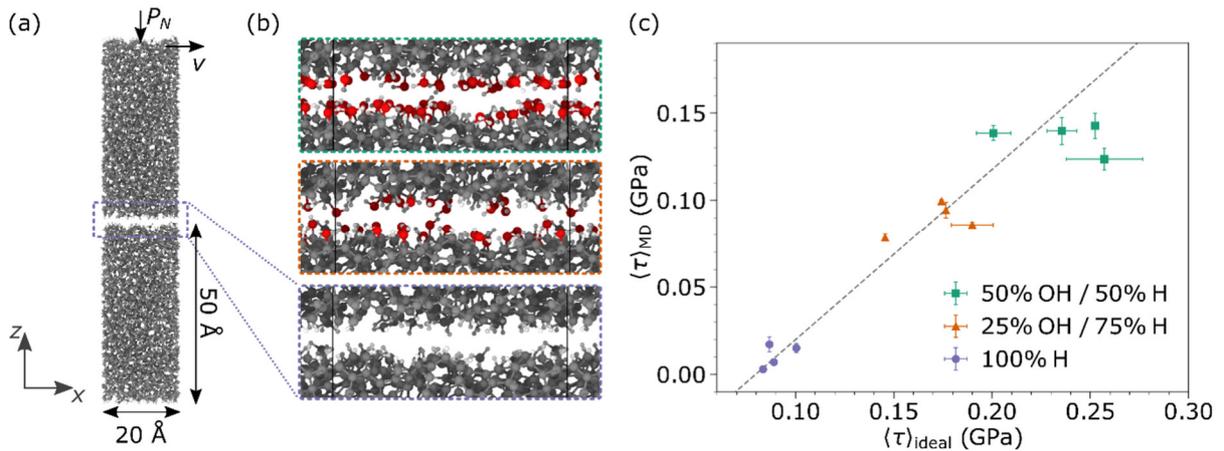

*Figure 14 (a) MD simulation setup of a dry sliding interface between two a-C slabs with a density of 2.5 g/cm$^3$ and 100% H termination. (b) Snapshots of the a-C interfaces with 50% OH / 50% H, 25% OH / 75% H, and 100% H surface terminations (from top to bottom). (c) Average shear stress $\langle\tau\rangle_{MD}$ for the surface pairings of panel (b) measured in MD sliding simulations as a function of $\langle\tau\rangle_{ideal}$, which is a measure for the corrugation of the CPES (see*

*text for details). For each of the three surface terminations 50% OH / 50% H (green squares), 25% OH / 75% H (orange triangles) and 100% H (purple circles), two different a-C samples and two different sliding directions are considered. The dashed grey line is a linear fit to the plotted data.*

## V.2 Relationship between friction and water film thickness

Another condition of considerable interest in boundary lubrication systems is one in which nanometric lubricant films are present at the contacts between surface asperities. In such cases, the relationship between friction and lubricant film thickness is particularly interesting as it does not necessarily follow continuum models. In this example, we consider model systems presented in the previous example (Section V.1), namely one model contact for each surface termination, and add different amounts of water molecules between the two contacting surfaces of each system. The amount of water ranges between 30 and 800 $H_2O$ molecules, corresponding to densities between 0.075 Å$^{-2}$ and 2 Å$^{-2}$ (Figure 15a). To determine $\langle\tau\rangle_{MD}$, we follow a protocol analogous to that described in Section V.1 and keep the $H_2O$ molecules rigid using the SHAKE algorithm[70] during the MD simulations. Upon sliding, the thickness of the water films $h_{H_2O}$ converges to a constant height that ranges between about 2 Å and 48 Å for the minimum and maximum amounts of water molecules, respectively (Figure 15a). The water film thickness is measured as the average difference between the maximum and the minimum $z$-coordinate of O atoms in $H_2O$ during the time interval from 0.5 ns to 2 ns.

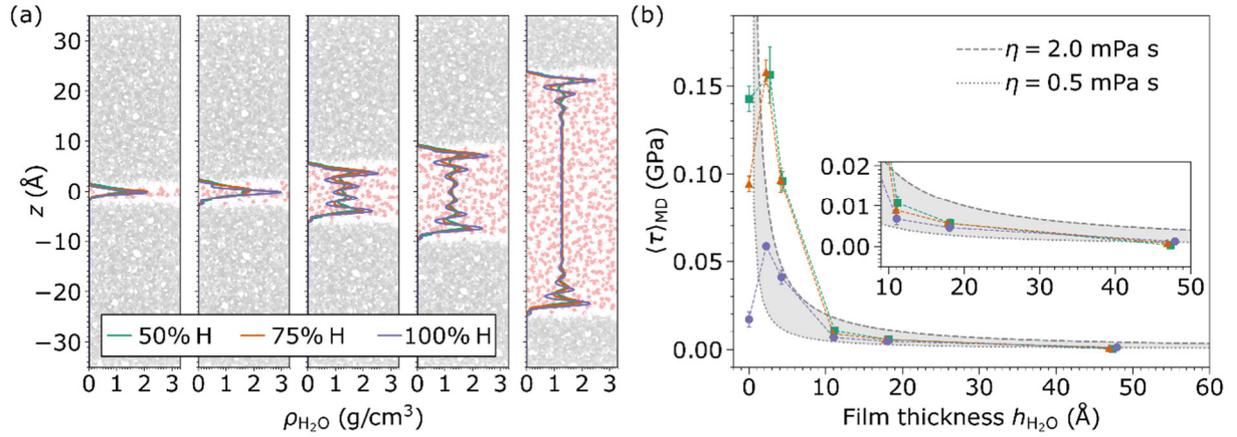

*Figure 15 (a) Snapshots of water-lubricated, 100% H-terminated a-C interfaces with different water film thicknesses $h_{H_2O}$ and averaged water density profiles $\rho_{H_2O}(z)$ for different surface termination ratios. The profiles are averaged over the time interval from 0.5 to 2 ns and are collected in slices with height 0.02 Å. $z = 0$ corresponds to half of the system height at the end of the respective simulation. (b) Dynamic shear stress $\langle\tau\rangle_{MD}$ as a function of $h_{H_2O}$ for a-C surfaces with different surface terminations (the colour code follows the legend in panel (a)). The inset shows a zoom-in for $h_{H_2O} \geq 9$ Å. Additionally, a continuum estimate of the shear stress $\tau = \eta v/h_{H_2O}$, where $v = 10$ m/s is the applied sliding velocity, is shown as the grey-shaded area for water viscosities $\eta$ between $0.5 \, mPa \, s$ (dotted grey line) and $2.0 \, mPa \, s$ (dashed grey line).*

Figure 15b shows $\langle\tau\rangle_{MD}$ as a function of $h_{H_2O}$, where $h_{H_2O} = 0$ corresponds to dry sliding. For each of the three surface termination ratios, the curves follow the same trend. Small traces of water lead to a significant increase of the shear stress compared to the dry contacts, with a maximum at 30 molecules, corresponding to a density of 0.075 Å$^{-2}$ (i.e., less than one water monolayer). With increasing film thickness, beyond the maximum, the shear stresses decrease monotonically. The maximum friction coefficients for the OH/H cases (estimated by $\frac{\langle\tau\rangle_{MD}}{P_N} \sim 0.16$) are approximately in line with the values ($\frac{\langle\tau\rangle_{MD}}{P_N} \sim 0.20$) determined with tight-binding simulations in Ref. 15 for OH/H-terminated diamond sliding interfaces that are lubricated with similar amounts of water molecules. Moreover, this shear-stress maximum is reminiscent of the non-monotonic dependence of friction on the adsorbate surface coverage observed in MD simulations of a tip/flat contact with a maximum friction observed at sub-monolayer adsorbate coverage[71].

In contrast to the dry case, the influence of the amount of OH surface terminations on $\langle\tau\rangle_{MD}$ disappears upon lubrication with minimal amounts of water. We only observe a difference between the hydrophilic OH-terminated cases and the hydrophobic 100% H-terminated case, for which significantly smaller friction is observed for $0 < h_{H_2O} < 10$ Å. For $h_{H_2O} > \sim 10$ Å, the shear stresses become independent of the termination type and the friction values (estimated by the ratio between $\langle\tau\rangle_{MD}$ and the applied normal load) become smaller than 0.01, i.e., the system enters the superlubricity regime[25].

In Figure 15b we additionally compare $\langle\tau\rangle_{MD}$ to simple estimates of the shear stresses for Newtonian fluids as obtained from $\tau = \eta v/h_{H_2O}$. Here, η is the viscosity of water that is assumed to be constant, $v$ is the applied sliding velocity and $v/h_{H_2O}$ approximates the average local shear rates. According to experimental results[72] at pressures in the GPa range and temperatures between 20°C-100°C, η should be in the range between 0.5 mPa s and 2.0 mPa s. We note that despite turning off the thermostat upon sliding, in our simulations the local temperatures at the interface remain below 380 K. This simple continuum approximation agrees remarkably well with the $\langle\tau\rangle_{MD}$ values from our MD simulations down to water thickness values as small as 11 Å. Unsurprisingly, for smaller $h_{H_2O}$, where the definition of the shear rate as $v/h_{H_2O}$ is no longer appropriate, $\langle\tau\rangle_{MD}$ starts to deviate from the continuum prediction, in particular for the H/OH-terminated cases.

Figure 15a also shows the water density profiles $\rho_{H_2O}$ along the z-direction (perpendicular to the interface plane). Independently of the a-C surface termination, the water molecules tend to arrange in layers parallel to the surfaces. In the systems with $h_{H_2O} \approx 2$ Å and $h_{H_2O} \approx 4$ Å, we observe a single layer of H$_2$O molecules. A markedly layered water structure with multiple distinct layers is visible for $h_{H_2O} \approx 11$ Å and with $h_{H_2O} \approx 18$ Å. In spite of this, the continuum description discussed above already gives a reasonable estimate of the local shear stress for these cases. Already at $h_{H_2O} \approx 18$ Å the density of a bulk water film is reached at the centre of the film. A bulk-like behaviour of the water molecules, with constant density $\rho_{H_2O}$ far from the surfaces, can be observed at $h_{H_2O} \approx 48$ Å.

### VI. Possible improvements and extensions of the force field

The development of the FF is based on a relatively simple and physically sound approach, which should prevent completely incorrect physical predictions in the model systems and application areas envisaged by the development. This is confirmed by the tests we have shown above, which consistently reveal an accurate description of numerous trends in forces and energies. In some cases, the accuracy of the FF in reproducing DFT test cases is even quantitative. These qualities make the FF a suitable tool for correlating changes in surface structure with changes in, for example, CPES and friction. However, our tests also indicate that there are situations in which the FF must be specifically validated or even further developed, mostly because of the limitations caused by its simple functional form. Moreover, there are surface termination types that have not been considered in the development yet and whose description requires further testing and, in some cases, relatively straightforward extensions of the parameter set. Similarly, the current version of the FF only incorporates water as a lubricant, but because the FF has been designed to be compatible with the OPLS FF, other lubricant molecules can be included in the FF rather easily. The following list provides an overview of some of these aspects that could be targeted in future extensions and developments and shows accuracy tests for the FF used in combination with existing OPLS parameters for glycerol.

- The FF is specifically tailored to a-C sliding interfaces and was developed using diamond (111) surfaces and graphite zigzag edges as reference systems. Since our parameterization is mainly based on the hybridisation of the carbon atom, it can most likely also be applied to other surface orientations (e.g., H/OH-passivated unreconstructed and 2x1-reconstructed (100) diamond surfaces or the aromatic Pandey-reconstructed (111) surface). However, specific tests should be carried out to test the FF accuracy in these cases.

- Quantum-mechanical simulations show that the dissociative chemisorption of molecules like $H_2$, $O_2$ and $H_2O$ at carbon tribological interfaces can result in a much richer variety of surface passivation species than just H atoms and hydroxyl groups (OH), including in particular carboxyl (C=O) and ether (C-O-C) surface groups[17,19,73]. Our results show that the original OPLS parameters for hydrocarbons and alcohols perform remarkably well for H/OH surface passivation groups. This suggests that the original OPLS parameters[34] for ethers and ketones should also be a good starting point to extend our FF to include these termination types. Similarly, standard OPLS parameters could serve as initial guesses to describe a-C surface termination by longer molecular fragments that can originate from the chemisorption of friction modifiers[21] or from tribologically induced surface oligomerization[13].

- Our tests show that the description of water adsorption on carbon surfaces is in some cases affected by the limitations of the TIP3P water model, particularly on hydrophobic surfaces, where the adsorption energies are systematically underestimated (e.g., Figure 9 and Figure 10). The attractive, non-bonded interaction between water molecules and carbon surfaces that are predominantly H-terminated is mostly determined by van der Waals interactions (in contrast to hydrophilic surfaces where Coulombic interactions are dominant)[41]. However, the TIP3P model only considers van der Waals interactions of water molecules via one LJ interaction site that is centred on the oxygen atom (i.e., the $\varepsilon$ parameters of H in $H_2O$ are zero). In spite of this, our tests show that the FF can reproduce the correct ranking of the $H_2O$ adsorption energies for different surface terminations. If needed, alternative water models (e.g., the CHARMM version of TIP3P that includes LJ sites on H[74], the TIP5P water model with additional Coulomb interaction sites[75], or the flexible SPC model, where the O–H bond length is not fixed[76]) could be tested in combination with this FF provided that a new FF parameter set is developed using the same DFT reference data and optimisation strategy presented in this work.

- Closely related to the previous point, for hydroxyl terminations on carbon surfaces we follow the convention of the original OPLS FF[34] and only use one LJ interaction site per OH group centred on the oxygen atom. This makes the Pauli repulsion of the OH group that is represented by the LJ term spherically symmetric. An additional LJ interaction site on the H atom could be used to improve the FF in this aspect. This, however, requires additional development and testing steps and could affect the compatibility with the OPLS FF.

- Non-bonded interactions between carbon atoms in graphitic layers are described in the FF by LJ interactions. However, the overlap of the π-orbitals between the graphitic layers is anisotropic and cannot be accurately described by a pair-potential[77]. As a result, the FF is not ideally suited to describe the potential energy corrugation between two graphene layers. This shortcoming is well known in the literature and could be accounted for by adding an additional stacking-dependent interlayer interaction as in the potential by Kolmogorov and Crespi[77]. Moreover, we also find that for extended graphite systems the out-of-plane bending of individual layers is barely penalised. This can be prevented by the additional inclusion of C–C–C–C dihedrals using the original OPLS parameters for benzene[55]. However, both the correct description of the out-of-plane bending and of the interlayer corrugation should not be critical for the modelling of aromatic surface patches on a-C that are intrinsically rough.

- The FF is designed to be compatible with the OPLS FF, which potentially allows the FF to be used for the description of a-C in contact with other liquids. In the context of tribology, a particularly relevant application case are a-C interfaces that are lubricated with hydrocarbons or alcohol-water mixtures. To test this compatibility, we select glycerol as a tribologically relevant example[5,6] and calculate rigid adsorption curves of a-C – glycerol systems analogously to the $H_2O$ adsorption calculations of Section III.5. We reuse the a-C systems from Section IV.3 with densities 2.0 g/cm³, 2.5 g/cm³ and 3.0 g/cm³ and consider one 100% H- and one 50% H-terminated case for each density. After an initial DFT relaxation of the physiosorbed glycerol, where both a-C and the molecule are not constrained, we keep both molecule and a-C rigid and evaluate the energy of the system as a function of their relative distance by means of DFT and the FF. Here, glycerol is described using the standard OPLS parameters[34], which are listed in Table S5, and for the glycerol – a-C interaction we use geometric-mean combination rules. Figure 16 shows the results of these calculations. The FF clearly captures the difference between hydrophilic and hydrophobic surfaces for all densities and the deviation of the FF and the DFT results in the energy minimum ranges from 0.03 eV to 0.24 eV, while the position of the minima differs by at most 0.13 Å. This indicates that indeed the FF can be used to describe a-C lubricated not only by water, but also by glycerol and potentially by other liquids included in the OPLS FF.

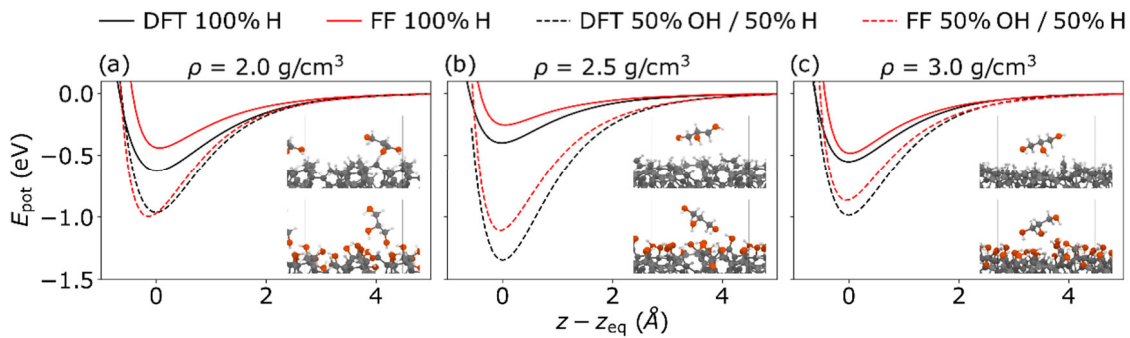

*Figure 16 Rigid adhesion curves of glycerol on 100% H- and 50% OH / 50% H-terminated a-C surfaces with densities of (a) 2.0 g/cm³, (b) 2.5 g/cm³ and (c) 3.0 g/cm³ as evaluated by DFT (black) and by the FF in combination with the standard OPLS description for glycerol[34] (red).*

*The insets show pictures of the corresponding structures in the respective DFT energy minimum.*

## VII. Conclusions

This article describes the development and validation of a DFT-based, non-reactive FF for dry and water-lubricated a-C interfaces that is specifically tailored for tribological simulations. It reproduces with good accuracy DFT results on solid-solid adhesion, water adsorption, a-C's elastic bulk properties, and the geometric and energetic corrugation of a-C/a-C and a-C/water interfaces. Its functional form closely follows the one of standard non-reactive FFs for liquids, and we demonstrate here that also crystalline and amorphous solids as well as their surfaces can be described accurately within this framework.

The FF is designed to be compatible with the OPLS FF, which facilitates future extensions and the description of additional lubricants and/or surface terminations can be based on already existing OPLS parameters as demonstrated using the example of glycerol. This applies not only to liquid lubricants, but also to solid lubricants such as PTFE[78] and graphite[79]. While we show that in many cases a further optimisation of the original OPLS parameters for hydrocarbons and alcohols can improve the description of the passivated a-C surfaces, the original parameters are sufficiently accurate in many of our tests. Yet, the newly developed FF adds a completely new aspect to traditional FFs for liquids, namely the possibility of using the same functional form with relatively simple parameters to accurately describe the elastic properties of the solid in contact with the liquid, in this case crystalline and amorphous carbon.

The FF enables dynamic simulations of chemically inert sliding interfaces consisting of hundreds of thousands of atoms for tens of nanoseconds. So far these length and time scales have been inaccessible for water/a-C systems because quantum-mechanical methods are usually necessary to accurately describe the complexity of solid/liquid interfaces made of C, O and H atoms, even in situations in which chemical reactions can be neglected. Hence, as briefly demonstrated in Section V, this FF can be used as an efficient tool to target a large set of scientific problems related to chemically passivated tribological interfaces. For example, extensive MD studies can be performed to study the relationship between friction and parameters such as sliding velocity, normal load, and the structure of the interfaces (e.g., atomic-scale roughness, material density, density and type of the surface passivation species, thickness of the lubricant film). Further goals could be the investigation of energy dissipation mechanisms in friction, e.g. via phonons[69] or, beyond tribology, of the wetting properties of a-C as a function of the surface chemical structure[41]. Regardless of the particular application, it is important to highlight that since the non-reactive FFs allow to seamlessly modify the individual, physically-motivated parameters such atomic charges, bond lengths or bond stiffnesses (see e.g. Refs. 21,24,37,80 for examples), also the influence of the individual parameters on friction becomes evaluable.

More in general, the development approach we described in this work can be helpful for an efficient development of interatomic interaction models for tribologically loaded interfaces that include a large variety of elements in different local chemical environments. Nonetheless, finding optimal parameter sets remains a tedious task and the development and application of automatised workflows[81,82] that use learning sets specifically tailored for tribology or consisting of pre-existing DFT results from publicly accessibly databases[83,84] is desirable. In fact, our 6-step optimisation procedure, that largely simplifies the optimisation problem by decoupling the optimisation of the individual parameters as far as possible, might serve as a suitable starting point for this purpose.

Acknowledgements

We thank Wolfgang Berger, Joachim Otschik, Ravindrakumar Bactavatchalou, Matthias Baldofski, Stefan Peeters and Takuya Kuwahara for fruitful discussions and Andreas Klemenz for help with the implementation of the FF. We gratefully acknowledge the Gauss Centre for Supercomputing e.V. for providing computing time through the John von Neumann Institute for Computing (NIC) on the GCS Supercomputer JUWELS[85] at Jülich Supercomputing Centre (JSC). Additional computational time was granted by the state of Baden-Württemberg through bwHPC and the DFG (grant no. INST 39/963-1 FUGG, bwForCluster NEMO). We acknowledge funding by the Fraunhofer Society through the PREPARE project SupraSlide and by the DFG within the Research Unit 5099.

Visualization of atomic structures was carried out with OVITO[86]. Input structures for the DFT and FF calculations were generated with ASE[87] and matscipy[52]. For the DFT calculations we used the Quickstep module of CP2K[50], FF calculations were carried out with LAMMPS[51]. For the density-functional tight-binding simulations we used the ASE[87] interface of Atomistica (https://github.com/Atomistica/atomistica). For the screened REBO2[65,68] and screened Tersoff[65,66] calculations we used the ASE[87]-Atomistica and the LAMMPS[51]-Atomistica interfaces.

## Supporting Information

Force-field parameters; Optimisation of Lennard-Jones parameters of $sp^2$-C atom types; Optimisation of charges of surface atom types; Original OPLS parameters; Electrostatic potential above crystalline surfaces/edges: error estimates for the fit set (75% H termination) and transferability test to 50% H-terminated crystalline surfaces/edges; Optimisation of the Lennard-Jones parameters of surface atom types; Relaxed adhesion and $H_2O$ adsorption configurations of crystalline carbon surfaces/edges; Protocols to generate a-C samples; Snapshots of relaxed adhesion and $H_2O$ adsorption configurations of a-C surfaces.

# An All-Atom Force Field for Dry and Water-Lubricated Carbon Tribological Interfaces

# Supporting Information

Thomas Reichenbach[1,†], Severin Sylla[1,2,†], Leonhard Mayrhofer[1], Pedro Antonio Romero[3], Paul Schwarz[3], Michael Moseler[1,2,4,5], Gianpietro Moras[1,*]

[1]*Fraunhofer IWM, MikroTribologie Centrum μTC, Wöhlerstraße 11, 79108 Freiburg, Germany*

[2]*Institute of Physics, University of Freiburg, Hermann-Herder-Straße 3, 79104 Freiburg, Germany*

[3]*Freudenberg Technology Innovation SE & Co. KG, Hoehnerweg 2-4, 69469 Weinheim, Germany*

[4]*Freiburg Materials Research Center, University of Freiburg, Stefan-Meier-Straße 21, 79104 Freiburg, Germany*

[5]*Cluster of Excellence livMatS @ FIT – Freiburg Center for Interactive Materials and Bioinspired Technologies, University of Freiburg, Georges-Köhler-Allee 105, 79110 Freiburg, Germany*

*Corresponding author: gianpietro.moras@iwm.fraunhofer.de

† Thomas Reichenbach and Severin Sylla contributed equally.

## S1. Force-field parameters

Tables S1-4 list the parameters of our optimised FF. An example of how to generate an interatomic bonding topology input file for LAMMPS[1] (i.e., the definition of the bonded interactions) and a corresponding FF parameter input file starting from a set of atomic coordinates and a list of all FF parameters is provided in the matscipy repository[2].

**Table S1**. LJ and point charge parameters (Eqs. 6 and 7 in the main text). Note that the LJ interaction for CG–CG and CG–CA atom pairs is truncated and shifted to zero at its minimum such that the interaction is purely repulsive. The atom types are labelled as in Table 1 in the main text.

| Atom type | $\varepsilon$ (meV) | $\sigma$ (Å) | $q$ (e) |
|---|---|---|---|
| CD | 0.000 | 0.000 | 0.0 |
| CB | 0.000 | 0.000 | 0.0 |
| CG | 2.906 | 3.340 | 0.0 |
| C1 | 7.150 | 3.801 | -0.06 |
| C2 | 7.150 | 3.801 | 0.18 |
| C3 | 7.150 | 3.801 | -0.12 |
| C4 | 7.150 | 3.801 | 0.12 |
| C5 | 7.150 | 3.801 | -0.18 |
| C6 | 7.150 | 3.801 | 0.06 |
| CZ | 4.309 | 3.729 | -0.115 |
| CY | 4.309 | 3.729 | 0.02 |
| CX | 4.309 | 3.729 | -0.23 |
| CA | 2.906 | 3.340 | 0.0 |
| O1 | 5.573 | 3.061 | -0.59 |
| O2 | 5.262 | 2.993 | -0.44 |
| H1 | 0.921 | 2.176 | 0.06 [3] |
| H2 | 1.626 | 2.120 | 0.115 [3] |
| H4 | 0.000 | 0.000 | 0.41 |
| H5 | 0.000 | 0.000 | 0.42 |
| OW | 4.423 [4] | 3.188 [4] | -0.83 [4] |
| HW | 0.000 [4] | 0.000 [4] | 0.415 [4] |

**Table S2**. Harmonic bond parameters for atom pairs $i$–$j$ (Eq. 3 in the main text). $sp^3$-C comprises the atom types CD, C1, C2, C3, C4, C5, C6, while $sp^2$-C includes CZ, CY, CX, CA, CB, CG. For the bonded parameters involving surface and bulk C atoms, we use the bulk parameters (e.g., C1–CD bond parameters are chosen according to the CD–CD bond parameters, also applies to graphite edges and a-C surfaces). Note that we have not yet considered CX bonded to other CX or to $sp^3$-C in the FF.

| Atom types | $k_r$ (eV/Å²) | $r_{eq}$ (Å) |
| --- | --- | --- |
| $sp^3$-C–$sp^3$-C | 15.614 | 1.543 |
| $sp^2$-C–$sp^2$-C | 21.121 | 1.426 |
| $sp^3$-C–$sp^2$-C | 18.367 | 1.484 |
| $sp^3$-C–H1 | 14.993 | 1.105 |
| $sp^2$-C–H2 | 17.218 | 1.091 |
| $sp^3$-C–O1 | 11.963 | 1.410 |
| CY–O2 | 18.437 | 1.365 |
| O1–H4 | 23.898 | 0.970 |
| O2–H5 | 23.727 | 0.976 |
| HW–OW | 19.514 [5] | 0.957 [6] |

**Table S3**. Harmonic angle parameters for $i$–$j$–$k$ three-body interactions (Eq. 4 in the main text). C comprises the atom types CD, C1, C2, C3, C4, C5, C6, CZ, CY, CX, CA, CB, CG.

| Atom types | $k_\theta$ (eV/rad²) | $\theta_{eq}$ (°) |
| --- | --- | --- |
| C–$sp^3$-C–C | 4.518 | 109.5 |
| C–$sp^2$-C–C | 3.957 | 120.0 |
| C–$sp^3$-C–H1 | 1.876 | 109.5 |
| C–$sp^3$-C–O1 | 2.004 | 109.5 |
| C–$sp^2$-C–H2 | 1.504 | 120.0 |
| C–CY–O2 | 2.715 | 120.0 |
| $sp^3$-C–O1–H4 | 2.243 | 109.112 |
| CY–O2–H5 | 2.591 | 110.222 |
| H1–$sp^3$-C–H1 | 1.876 | 109.5 |
| H1–$sp^3$-C–O1 | 1.876 | 109.5 |
| H2–CX–H2 | 1.504 | 120.0 |

| HW–OW–HW | 2.385 [5] | 104.52 [6] |

**Table S4**. Fourier coefficients for the four-body dihedral interactions $i$–$j$–$k$–$l$ (Eq. 5 in the main text).

| Atom types | $k_{\Phi,\gamma,1}$ (meV) | $k_{\Phi,\gamma,2}$ (meV) | $k_{\Phi,\gamma,3}$ (meV) | Reference |
|---|---|---|---|---|
| C–C–sp²-C–H2 | 0 | 459.7 | 0 | benzene [7] |
| H2–sp²-C–CZ–H2 | 0 | 459.7 | 0 | benzene [7] |
| O–C–sp²-C–H2 | 0 | 459.7 | 0 | benzene [7] |
| C–CY–O2–H5 | 0 | 72.9 | 0 | phenol [3] |
| C–sp³-C–O1–H4 | -15.4 | -7.5 | 21.3 | alcohol [3] |
| H1–sp³-C–O1–H4 | 0 | 0 | 19.5 | alcohol [3] |

## S2. Optimisation of Lennard-Jones parameters of sp²-C atom types

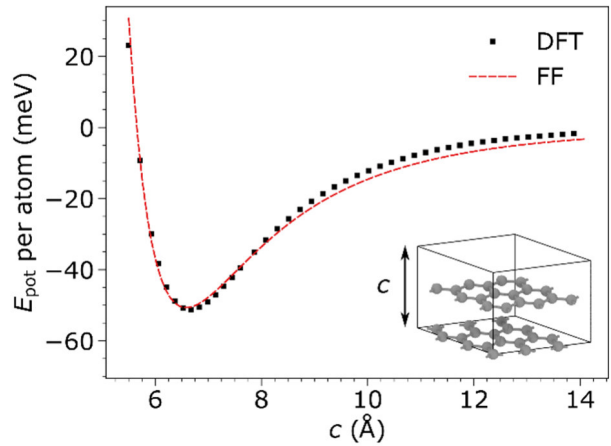

**Fig. S1.** Graphite energy per atom as a function of the lattice constant $c$ in the direction normal to the graphite layers as obtained by DFT (black squares) and the FF with optimised parameters (red dashed line). See Section III.2 in the main text for details.

## S3. Optimisation of charges of surface atom types

The charge parameters of C, O, and H surface atoms are optimised to fit the ESP above diamond surfaces and graphite edges with 75% H terminations. To fit the parameters, we use the REPEAT method[8] as implemented in CP2K[9] and the objective function

$$\Xi_{\text{REPEAT}}(Q) = \frac{1}{N}\sum_{k}^{N}\left(V_{\text{DFT}}(\vec{r_k}) - V_{\text{ESP}}(\vec{r_k}; Q) - \delta(Q)\right)^2,$$

where $V_{\text{DFT}}$ is the DFT ESP and $V_{\text{ESP}}$ the ESP generated by the point-charge model with a set of charges $Q = (q_1, q_2, \ldots)$. $V_{\text{ESP}}$ is shifted by $\delta(Q) = \frac{1}{N}\sum_{k}^{N}\left(V_{\text{DFT}}(\vec{r_k}) - V_{\text{ESP}}(\vec{r_k}; Q)\right)$ to ensure that the average values of the two potentials are aligned. Here, $N$ refers to the number of fit points $\vec{r_k}$. To evaluate the objective function, we follow the CP2K implementation and sample a region of vacuum that is defined as the union of cubic boxes with a 3 Å edge length positioned 2 Å away from each H and O surface atom along the $z$ direction. Within this region, the fit points $\vec{r_k}$ are chosen according to a cubic grid with a grid spacing of 0.1 Å.

## S4. Original OPLS parameters

Tables S5-S7 contain the original OPLS parameters for surface atom types that we use for comparison with our optimised parameter set. Since there are no original OPLS parameters for bulk C atom types, we use our optimised bulk parameters instead (Section S1) and apply the conventions described in the main text for mixed (bulk/surface, $sp^2/sp^3$) bonds and angles. The parameters for atom types OW and HW as well as the Fourier coefficients for the dihedral interactions are the same as in Section S1.

**Table S5**. Original OPLS LJ and point charge parameters. The atom types are labelled as in Table 1 in the main text.

| Atom type | $\varepsilon$ (meV) | $\sigma$ (Å) | $q$ (e) | Reference |
|---|---|---|---|---|
| C1 | 2.862 | 3.500 | -0.06 | $R_3CH$[3] |
| C2 | 2.862 | 3.500 | 0.265 | $R_3COH$[3] |
| C3 | 2.862 | 3.500 | -0.12 | $R_2CH_2$[3] |
| C4 | 2.862 | 3.500 | 0.205 | $R_2CHOH$[3] |
| C5 | 2.862 | 3.500 | -0.18 | RCH$_3$[3] |
| C6 | 2.862 | 3.500 | 0.145 | RCH$_2$OH[3] |
| CZ | 3.035 | 3.550 | -0.115 | benzene[3] |
| CY | 3.035 | 3.550 | 0.150 | phenol[3] |
| CX | 3.296 | 3.550 | -0.23 | H$_2$C=[3] |
| CA | 3.296 | 3.550 | 0.0 | R$_2$C=[3] |
| O1 | 7.372 | 3.120 | -0.683 | ROH[3] |
| O2 | 7.372 | 3.070 | -0.585 | phenol[3] |
| H1 | 1.301 | 2.500 | 0.06 | alkanes[3] |
| H2 | 1.301 | 2.420 | 0.115 | benzene[3] |
| H4 | 0.000 | 0.000 | 0.418 | ROH[3] |
| H5 | 0.000 | 0.000 | 0.435 | phenol[3] |

**Table S6**. Original OPLS harmonic bond parameters. In this table, sp$^3$-C comprises the atom types C1, C2, C3, C4, C5, C6, while sp$^2$-C includes CZ, CY, CX. For the bonded parameters involving surface and bulk C atoms, we use the fitted bulk parameters (e.g., C1–CD bond parameters are chosen according to the CD–CD bond parameters, also applies to graphite edges and a-C surfaces).

| Atom types | $k_r$ (eV/Å$^2$) | $r_{eq}$ (Å) | Reference |
|---|---|---|---|
| sp$^3$-C–sp$^3$-C | 13.443 | 1.526 | 7 |
| sp$^2$-C–sp$^2$-C | 20.337 | 1.400 | 10 |
| sp$^3$-C–H1 | 14.353 | 1.090 | 7 |
| sp$^2$-C–H2 | 14.744 | 1.080 | 7 |
| sp$^3$-C–O1 | 13.876 | 1.410 | 7 |
| CY–O2 | 19.514 | 1.364 | 10 |
| O1–H4 | 23.980 | 0.960 | 7 |
| O2–H5 | 23.980 | 0.960 | 7 |

**Table S7**. Original OPLS harmonic angle parameters. Here, C comprises the atom types CD, C1, C2, C3, C4, C5, C6, CZ, CY, CX, CA, CB, CG. For sp$^3$-C and sp$^2$-C the definition from Table S6 applies. The C–sp$^3$-C–C and C–sp$^2$-C–C angle parameters are our optimised bulk parameters from Table S3.

| Atom types | $k_\theta$ (eV/rad²) | $\theta_{eq}$ (°) | Reference |
|---|---|---|---|
| C–sp³-C–C | 4.518 | 109.5 | This work |
| C–sp²-C–C | 3.957 | 120.0 | This work |
| C–sp³-C–H1 | 1.518 | 109.5 | 7 |
| C–sp³-C–O1 | 2.168 | 109.5 | 7 |
| C–sp²-C–H2 | 1.518 | 120.0 | 7 |
| C–CY–O2 | 3.035 | 120.0 | 10 |
| sp³-C–O1–H4 | 2.385 | 108.5 | 10 |
| CY–O2–H5 | 1.518 | 113.0 | 10 |
| H1–sp³-C–H1 | 1.518 | 109.5 | 7 |
| H1–sp³-C–O1 | 1.518 | 109.5 | 7 |
| H2–CX–H2 | 1.518 | 120.0 | 7 |

## S5. Electrostatic potential above crystalline surfaces/edges: error estimates for the fit set (75% H termination) and transferability test to 50% H-terminated crystalline surfaces/edges

As a measure of how accurately the variance of the DFT ESP is reproduced by a set of classical point charges $Q$, we calculate the RRMS error as [8,11]

$$\text{RRMS}(Q) = \sqrt{\frac{\sum_k^N \left(V_{\text{DFT}}(\vec{r_k}) - V_{\text{ESP}}(\vec{r_k}; Q) - \delta(Q)\right)^2}{\sum_k^N V_{\text{DFT}}(\vec{r_k})^2}},$$

where $V_{\text{DFT}}$ is the DFT ESP, $V_{\text{ESP}}$ the Coulomb potential generated by the point charge model with a set of charges $Q = (q_1, q_2, ...)$ and $\delta(Q) = \frac{1}{N}\sum_k^N \left(V_{\text{DFT}}(\vec{r_k}) - V_{\text{ESP}}(\vec{r_k}; Q)\right)$. $N$ refers to the number of sampling points $\vec{r_k}$. Since the reference state of the ESP in a periodic system is not well-defined, the RRMS only allows for a relative comparison of the accuracy of two sets of partial charges.

In Table S8, the RRMS values are given for the fit set consisting of 75% H-terminated crystalline surfaces/edges, where $V_{\text{ESP}}$ is calculated with our optimised charges and with the original OPLS charges. The RRMS values are also shown for the ESP above 50% H-terminated surfaces/edges as a transferability test for the charge parameters. The sampled region is chosen as described in Section S3. In each of the considered cases, the RRMS value calculated with our optimised charges is lower than the one calculated with the original OPLS charges.

**Table S8**. RRMS values for the fit set with 25% OH / 75% H terminations and for test surfaces/edges (50% OH / 50% H termination) as obtained using the FF with optimised parameters (FF) and the original OPLS parameters for hydrocarbons and alcohols[3].

| Surface type | Parameter set | RRMS (%) |
|---|---|---|
| Diamond, 25% OH / 75% H termination | FF | 0.18 |
|  | OPLS | 0.34 |
| Graphite, 25% OH / 75% H termination | FF | 0.38 |
|  | OPLS | 0.86 |
| Diamond, 50% OH / 50% H termination | FF | 0.20 |
|  | OPLS | 0.32 |
| Graphite, 50% OH / 50% H termination | FF | 1.25 |
|  | OPLS | 1.77 |

## S6. Optimisation of the Lennard-Jones parameters of surface atom types

The objective function for the optimization of a set of LJ parameters $P = (\varepsilon_1, \sigma_1, \varepsilon_2, \sigma_2, ...)$ in Section III.5 reads

$$\Xi_{\mathrm{LJ}}(P) = \sqrt{\sum_i \sum_{k_i} W_i \big(E_{k_i,\mathrm{FF}}(P) - E_{k_i,\mathrm{DFT}}\big)^2},$$

where the summation index $i$ runs over the different adhesion and adsorption curves in the fit set (Figs. S2-S5) and the index $k_i$ runs over the different distances $z - z_{\mathrm{eq}}$ of each curve. The $i$-th term in the summation is normalised by the expression

$$W_i = \frac{w_i / (\Delta E_{i,\mathrm{DFT}})^2}{\sum_j w_j / (\Delta E_{j,\mathrm{DFT}})^2},$$

where $\Delta E_{i,\mathrm{DFT}}$ is the potential depth of the $i$-th adhesion/adsorption curve. The summation index $j$ runs over all different adhesion and adsorption curves in the fit set. The parameters $w_i$ are adjusted manually to tune the weights of the different curves and are listed in Table S9.

**Table S9**. Number of adhesion and adsorption curves used in the fit set for the LJ parameter optimization, along with the weights $w_i$. The weights are ordered according to their appearance in the Figs. S2, S3, S4 and S5.

| Atom types | Diamond surfaces | Graphite surfaces |
|---|---|---|
| Atom types for $\varepsilon, \sigma$ optimization | C1, H1, O1 | CZ, H2, O2 |
| No. of adhesion configurations | 4 | 4 |
| No. of $H_2O$ adsorption configurations | 2 | 3 |
| Weights | (1, 4, 20, 4, 10, 4) | (1, 10, 10, 1, 1, 1, 1) |

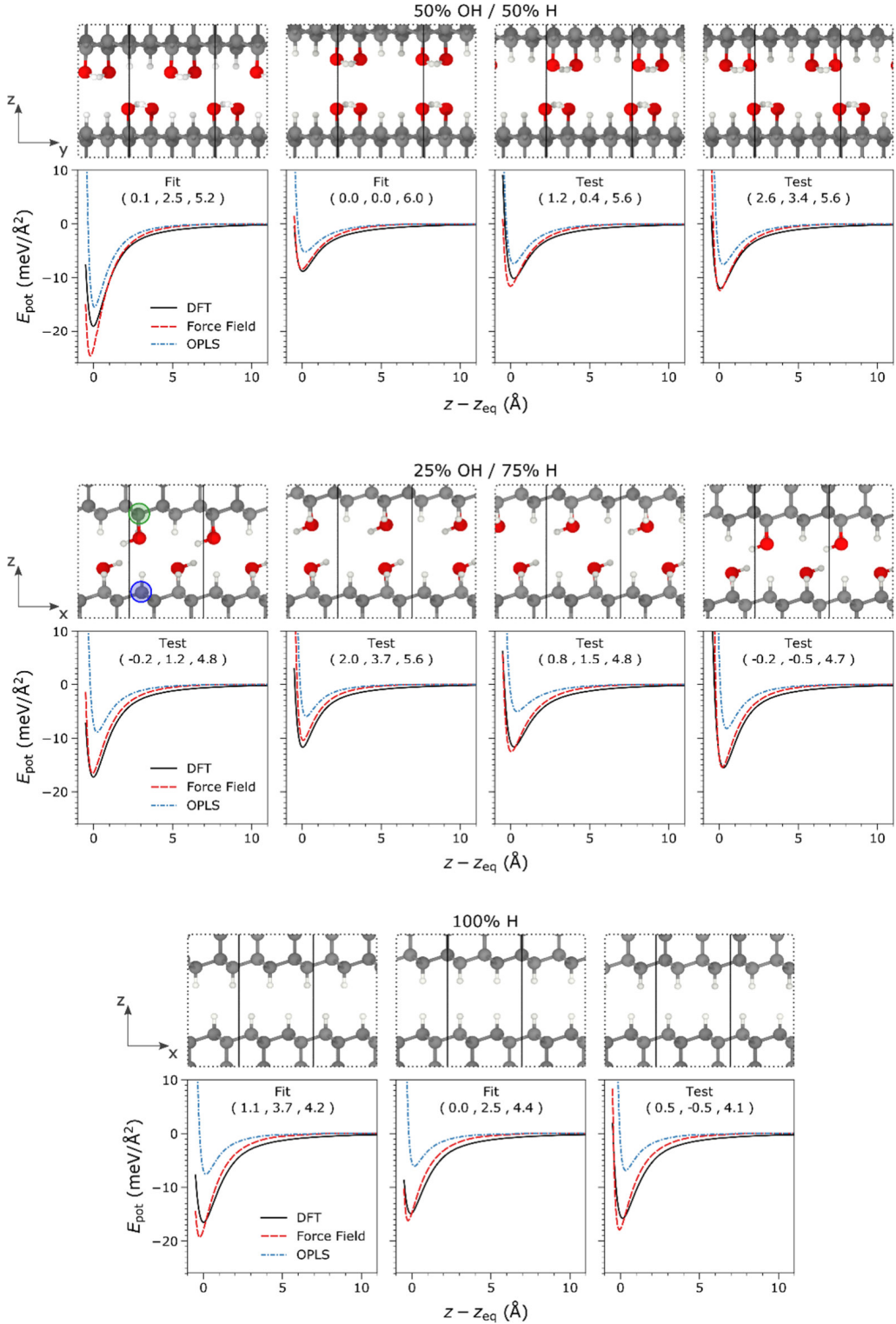

**Fig. S2.** Rigid diamond adhesion curves computed with DFT (black solid line) and with the FF with optimised parameters (red dashed line) and with original OPLS parameters (blue dash-dotted line). The energies are given per interface area. Snapshots of the corresponding structures in the DFT energy minimum are shown above each diagram. In each diagram it is indicated whether the data belongs to the fit or the test set. The 3-tuple in each of the diagrams is the distance vector (Å) between the two carbon atoms that are marked in green and blue in the first snapshot in the second row to indicate the lateral position between the upper and lower slab.

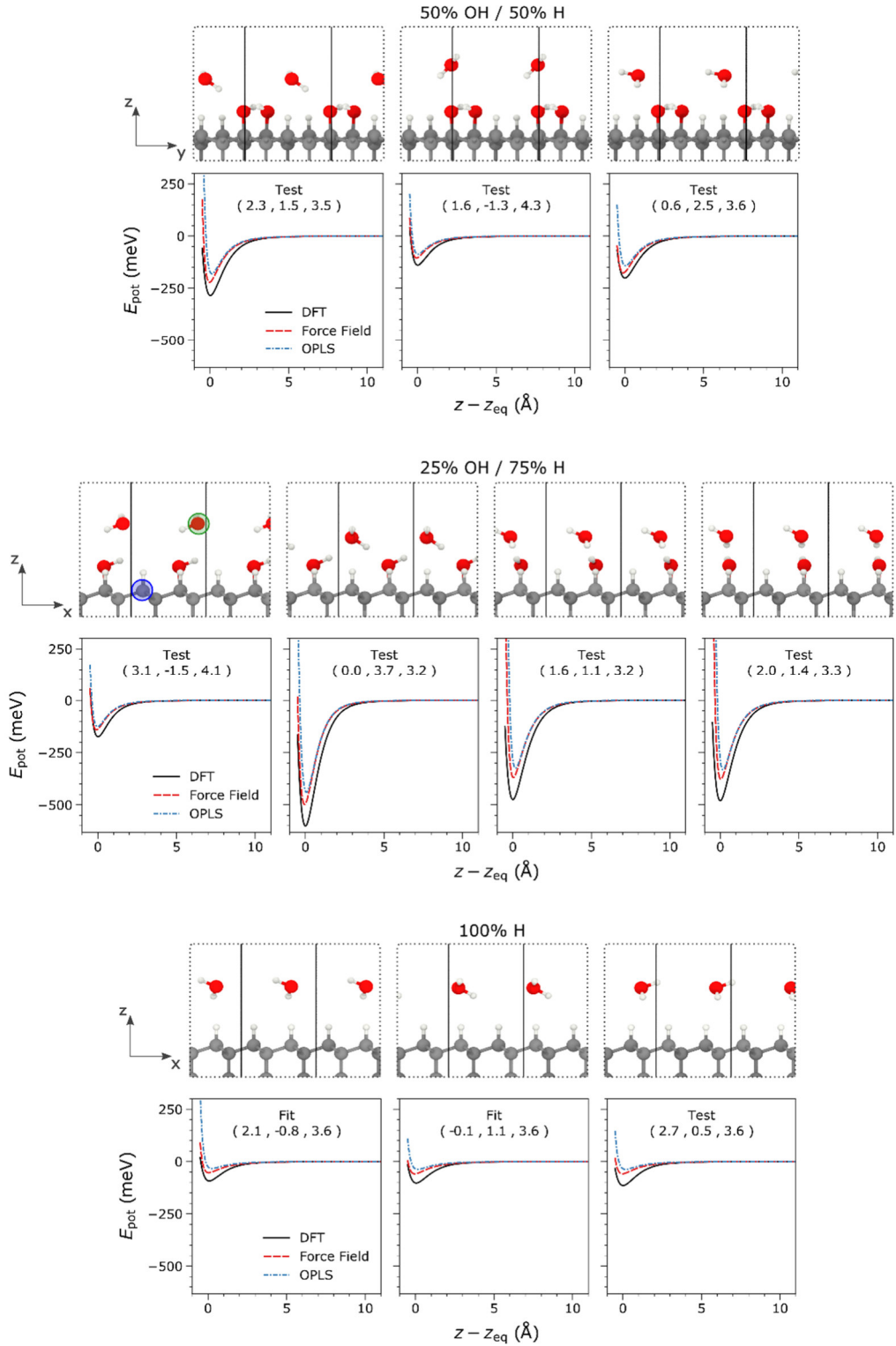

**Fig. S3.** Rigid diamond - H$_2$O adsorption curves computed with DFT (black solid lines) and with the FF with optimised parameters (red dashed lines) and with original OPLS parameters (blue dash-dotted lines). The energies are given per H$_2$O molecule. Snapshots of the corresponding structures in the DFT energy minimum are shown above each diagram. In each diagram it is indicated whether the data belongs to the fit or the test set. The 3-tuple in each of the diagrams is the distance vector (Å) between the two atoms that are marked in green and blue in the first snapshot of the second row.

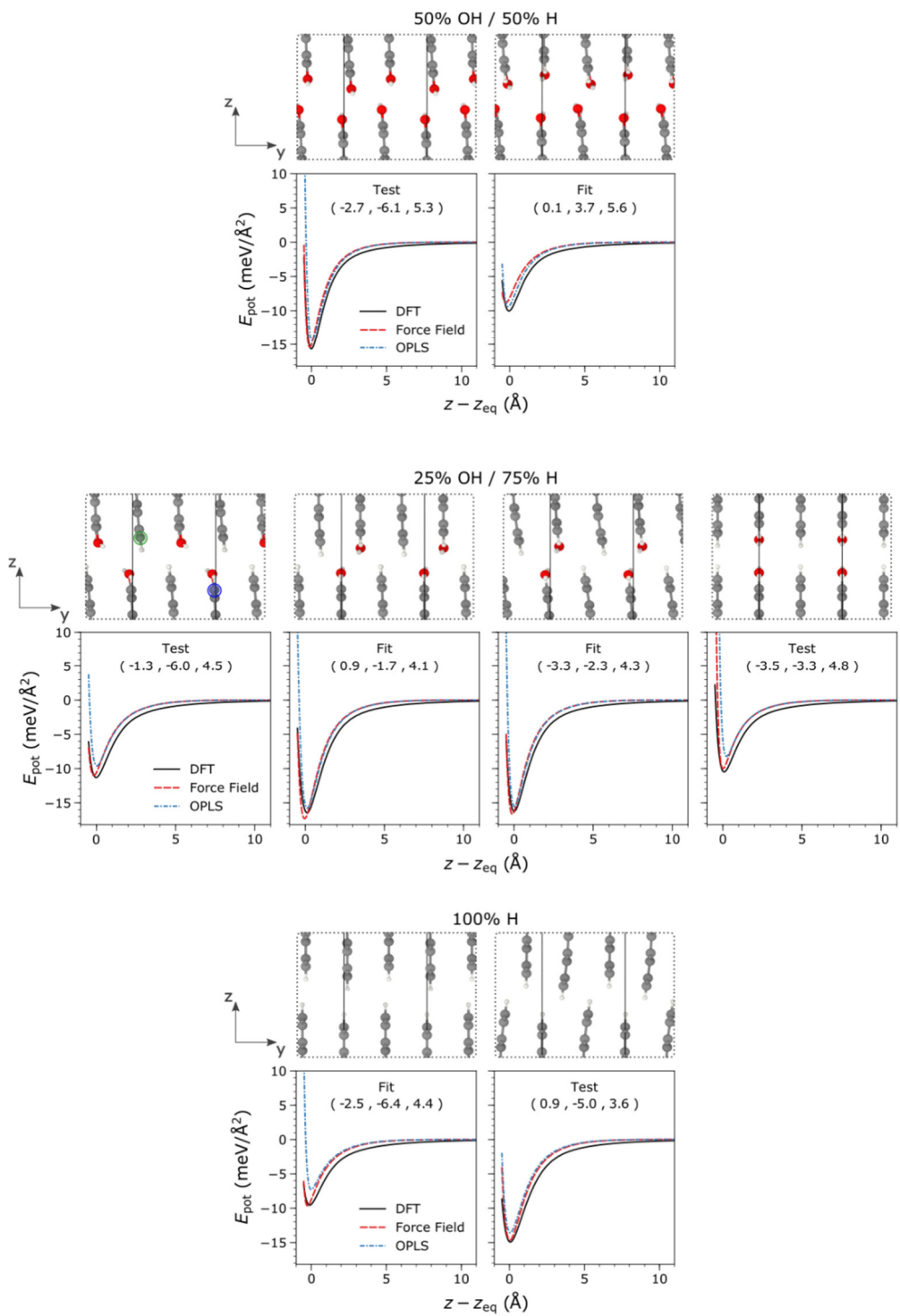

**Fig. S4.** Rigid graphite adhesion curves computed with DFT (black solid line) and with the FF with optimised parameters (red dashed line) and with original OPLS parameters (blue dash-dotted line). The energies are given per interface area. Snapshots of the corresponding structures in the DFT energy minimum are shown above each diagram. In each diagram it is indicated whether the data belongs to the fit or the test set. The 3-tuple in each of the diagrams is the distance vector (Å) between the two carbon atoms that are marked in green and blue in the first snapshot of the second row.

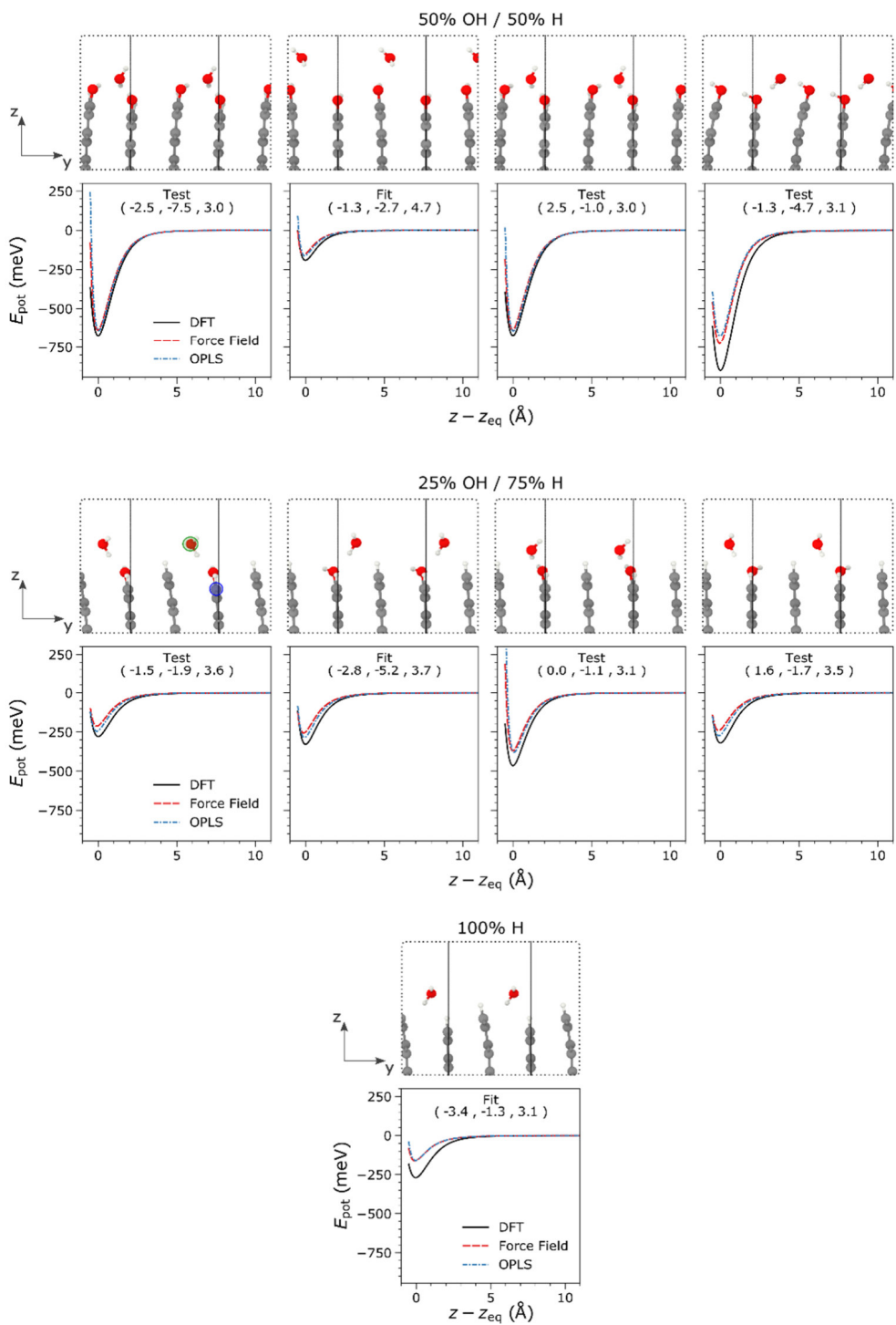

**Fig. S5.** Rigid graphite - $H_2O$ adsorption curves computed with DFT (black solid line) and with the FF with optimised parameters (red dashed lines) and with original OPLS parameters (blue dash-dotted lines). The energies are given per $H_2O$ molecule. Snapshots of the corresponding structures in DFT energy minimum are shown above each diagram. In each diagram it is indicated whether the data belongs to the fit or the test set. The 3-tuple in each of the diagrams is the distance vector (Å) between the two atoms that are marked in green and blue in the first snapshot of the second row.

## S7. Relaxed adhesion and H$_2$O adsorption configurations of crystalline carbon surfaces/edges

Figures S6-S9 show snapshots of the interface adhesion and H$_2$O adsorption tests with relaxed geometries for the crystalline carbon surfaces/edges (Section IV.1 in the main text).

For the calculation of the interface adhesion energies, in each relaxation, we fix the positions of the outermost atoms of the lower slab (within a thickness of 3 Å), and the $x$ and $y$ components of the positions of the outermost atoms of the upper slab (also within a thickness of 3 Å). In addition, for the adhesion of graphite edges, we also keep the $z$ component of the atomic positions of the outermost upper atoms fixed to avoid lateral displacements of the individual graphite layers. In this case, the distance between the outmost rigid layers is initially optimised using the FF. In the subsequent DFT relaxation and FF relaxation with original OPLS parameters the same distance is used. In the calculation of the water adsorption energies on these surfaces/edges, we keep the positions of the atoms within the lowest 3 Å of the slab fixed, while the H$_2$O molecules that are placed on top of the surface/edge remain unconstrained.

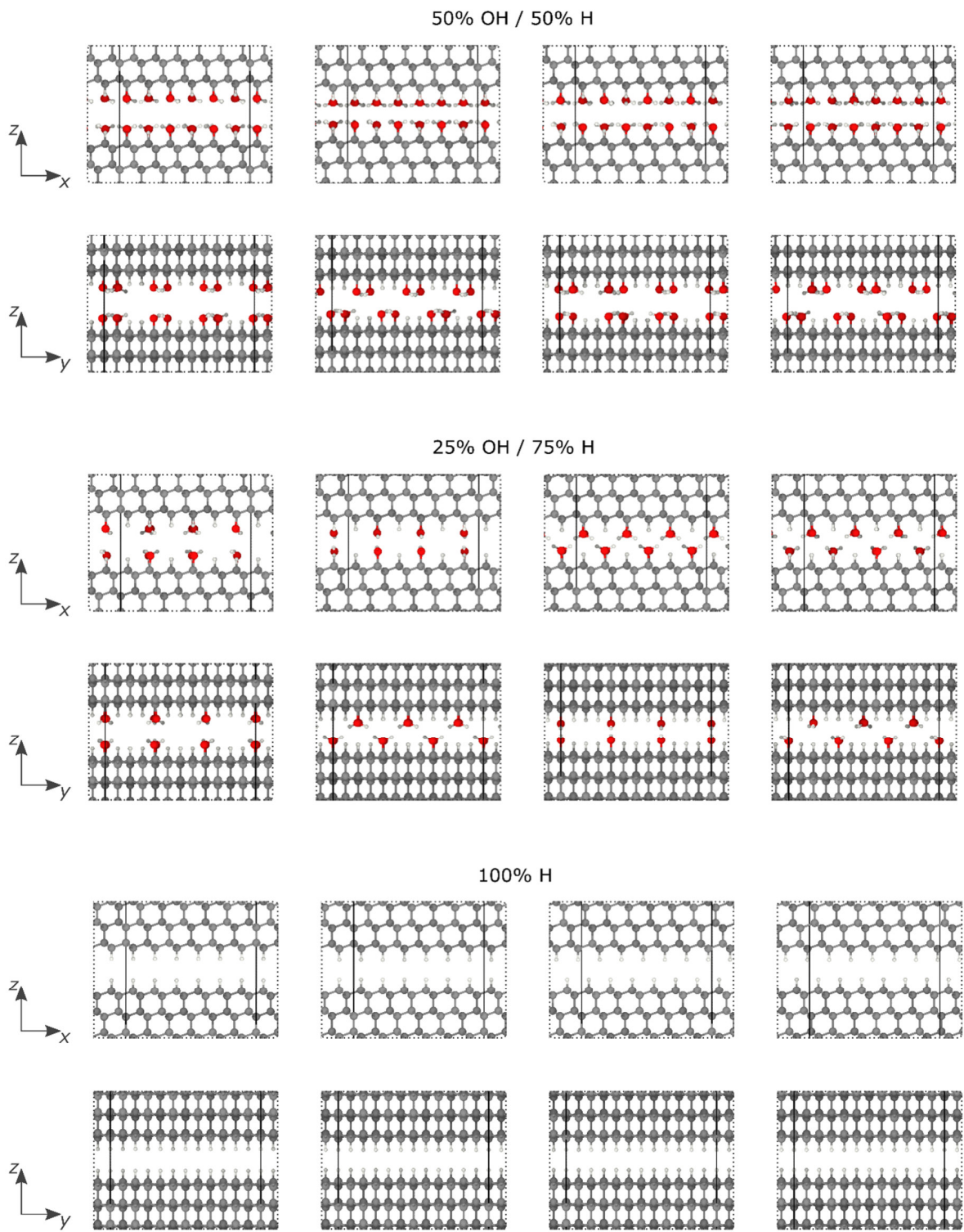

**Fig. S6.** Snapshots of the DFT-relaxed configurations used to calculate diamond adhesion energies (Fig. 9 in the main text). The columns show different lateral positions of the two slabs.

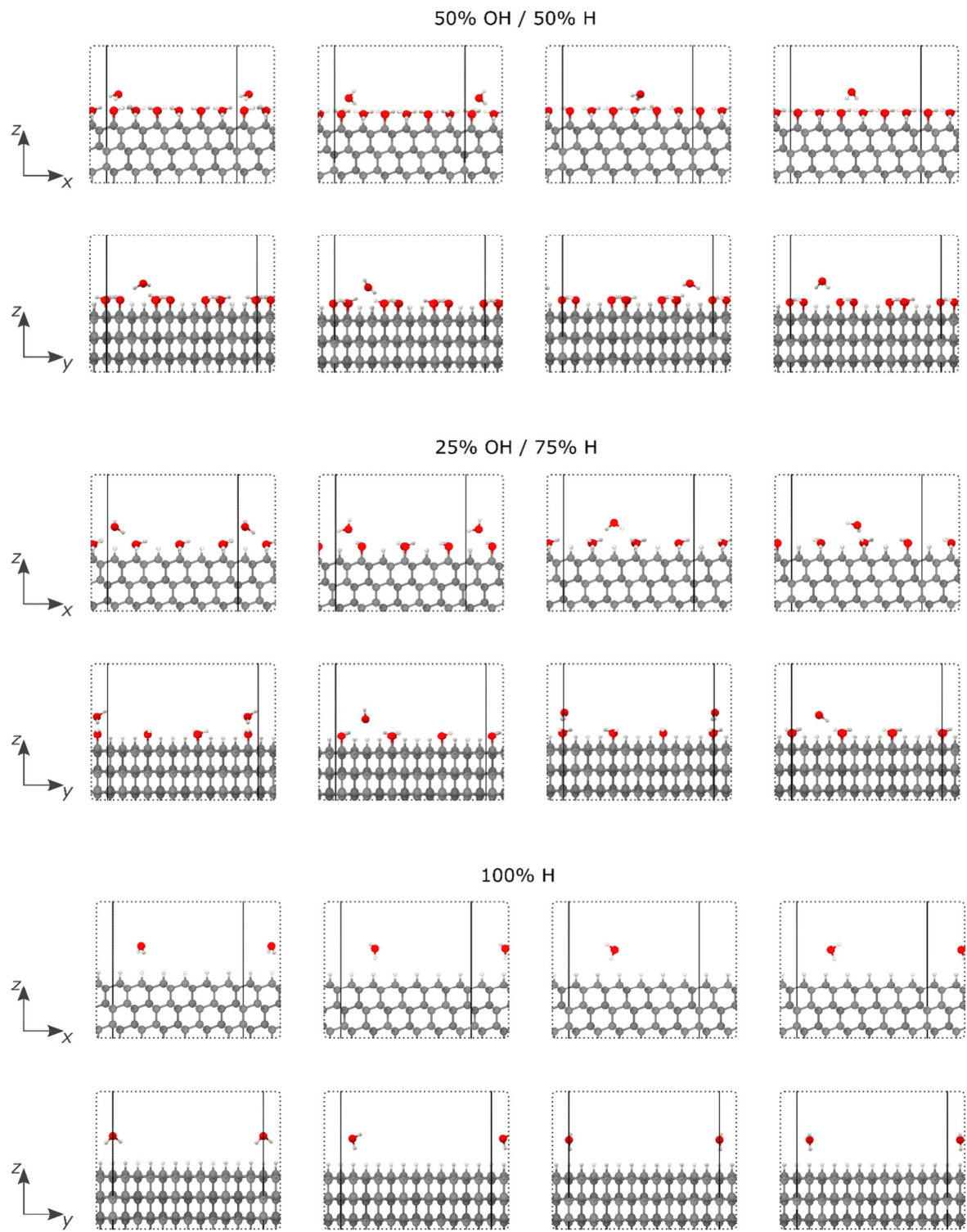

**Fig. S7.** Snapshots of DFT-relaxed configurations used to calculate the $H_2O$ adsorption energies on diamond (Fig. 9 in the main text). The columns show the relaxed configurations corresponding to different lateral positions of the $H_2O$ molecule.

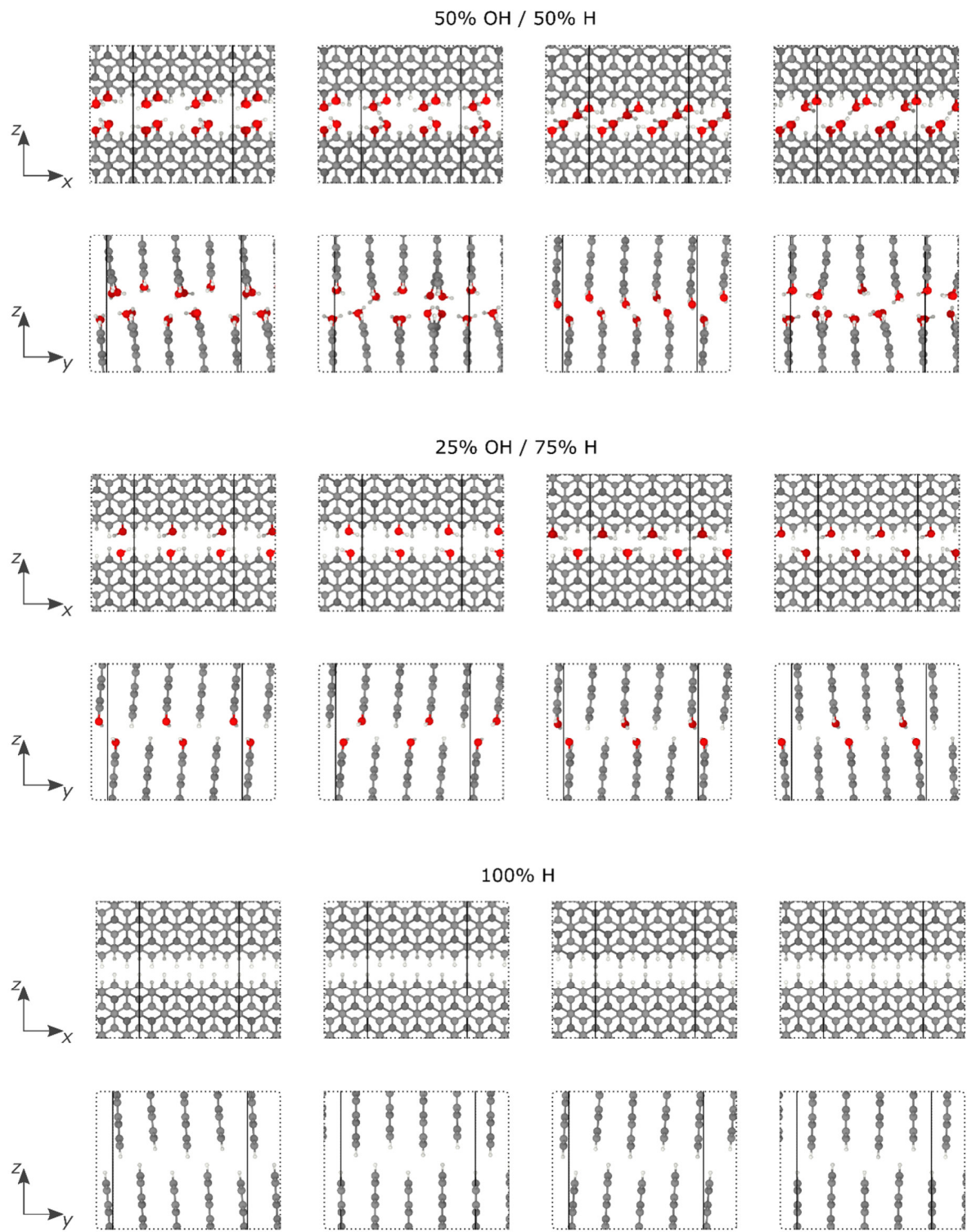

**Fig. S8.** Snapshots of DFT-relaxed configurations used to calculate the graphite adhesion energies (Fig. 9 in the main text). The columns show different lateral positions of the two slabs.

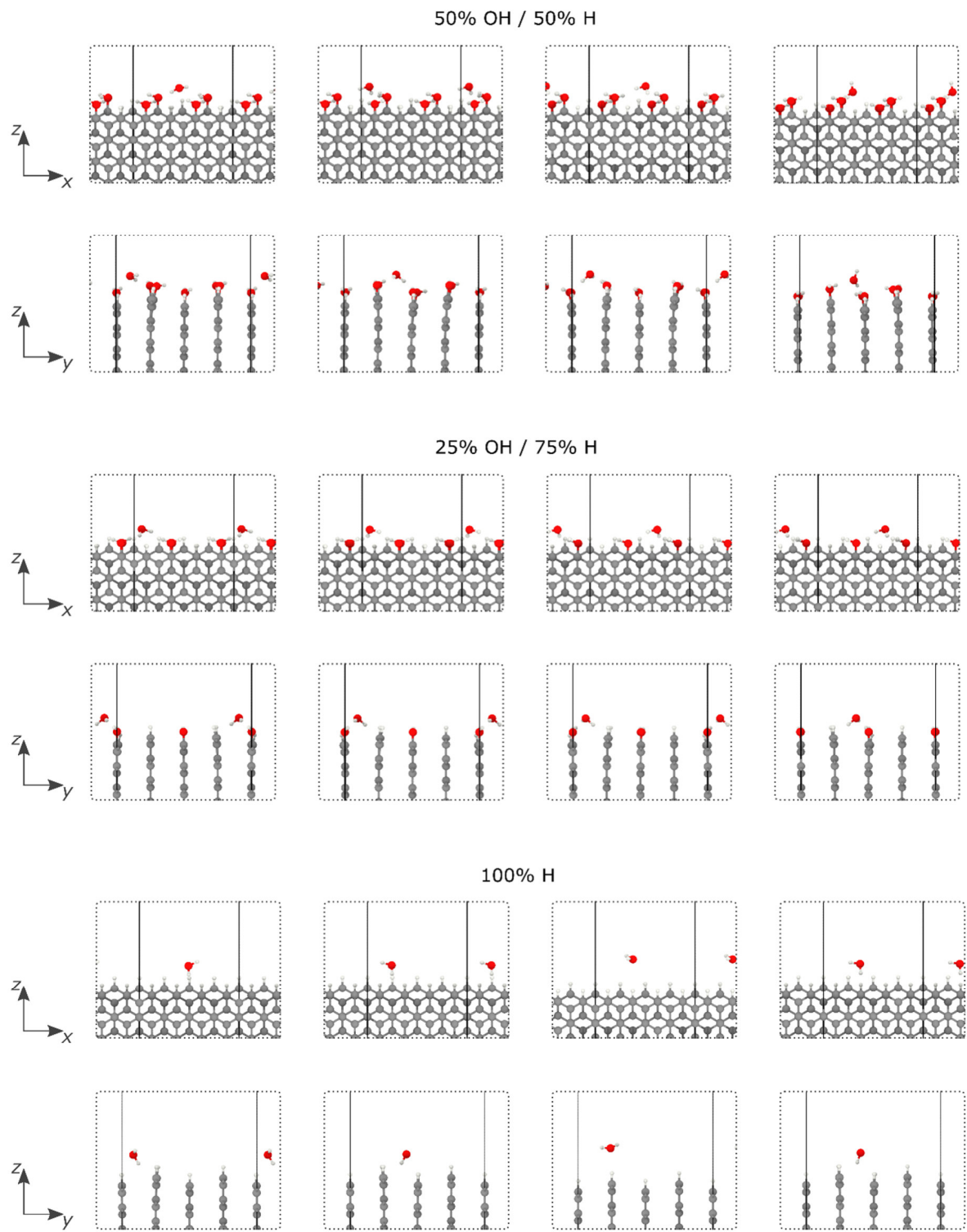

**Fig. S9.** Snapshots of DFT-relaxed configurations used to calculate H$_2$O adsorption energies on graphite edges (Fig. 9 in the main text). The columns show the relaxed configurations corresponding to different lateral positions of the H$_2$O molecule.

## S8. Protocols to generate a-C samples

### S8.1 Test structures for the calculation of elastic moduli, adhesion energies and H$_2$O adsorption energies

To generate bulk a-C samples of varying densities, we rapidly quench a liquid carbon phase using the protocol by R. Jana et al.[12]. Interatomic forces are modelled using the screened version[13] of the Tersoff potential[14]. To control the temperature of the system, we use a Langevin thermostat with a damping constant of 0.1 ps. Initially, carbon atoms are placed randomly inside a cubic simulation cell with edge length 12.257 Å. The number of atoms in each sample is between 138 and 300, corresponding to initial densities ranging from 1.50 g/cm$^3$ to 3.25 g/cm$^3$. Each system is equilibrated at 10000 K for 100 ps. Subsequently, it is quenched to 0 K at constant volume with a constant rate of 10 K/ps. Afterwards, we switch to DFT and relax the atomic positions and the simulation cell with a threshold of 0.01 eV/Å for the forces on each atom. During the last step, triclinic cell deformations are allowed.

For the densities 2.0 g/cm$^3$, 2.5 g/cm$^3$ and 3.0 g/cm$^3$, we create surface models by relaxing the quenched bulk structures (without cell relaxation), while releasing the periodic boundary conditions in the $z$-direction. The relaxation is performed using the screened Tersoff potential with a force threshold of 0.01 eV/Å. To terminate dangling bonds of the upper and lower surface with H atoms, we follow the acceptance-rejection method proposed in Ref. 15. We define the surfaces arbitrarily as regions with thickness of 2 Å starting from the atoms with maximal and minimal $z$-coordinate, respectively. During this procedure, we use SCC-DFTB[16] to calculate interatomic forces. Individual H atoms are iteratively placed in the vicinity of random undercoordinated surface atoms. In each step the system is relaxed with a force threshold of 0.025 eV/Å for the Euclidean norm of the force vector on each individual atom and the termination is accepted if the hydrogen adsorption energy is energetically favourable using H$_2$ molecules as a reservoir. After every undercoordinated atom has been tested in this way, we switch again to DFT and relax the final structure using a force threshold of 0.01 eV/Å for the Euclidean norm of the DFT force vector on each individual atom. Surfaces with mixed OH and H terminations (50% H or 75% H) are generated based on the fully H-terminated structures by replacing randomly selected H atoms by OH groups. While doing this, at most one OH group per carbon atom is allowed in order to limit the complexity of the systems. The resulting structures are again relaxed with a force threshold of 0.01eV/Å.

**S8.2 a-C structures for dry and water-lubricated sliding simulations**

To generate the samples for dry and water-lubricated sliding simulations (Section V in the main text) we follow the procedure described in Section S8.1, with some modifications due to the larger system size. We consider bulk a-C samples with a density of 2.5 g/cm$^3$ which consist of 2507 C atoms in an orthogonal simulation cell with lengths of 2 nm in the $x$- and $y$-directions and of 5 nm in the $z$-direction. Due to the high computational cost, we omit the DFT relaxation of the slabs after quenching. Moreover, during the termination procedure with H atoms, we use the screened[13] REBO2 potential[17] instead of SCC-DFTB. Mixed OH/H-terminated surfaces are again generated by replacing randomly selected H terminations by OH, but this time without any final DFT relaxation. We note that in case of these REBO2-relaxed a-C structures (as opposed to the structures from Section S8.1), there are occasionally sp-hybridised C atoms on the surface. For simplicity, we use the FF parameters for surface sp$^2$-C atoms to describe these atoms.

## S9. Snapshots of relaxed adhesion and H₂O adsorption configurations of a-C surfaces

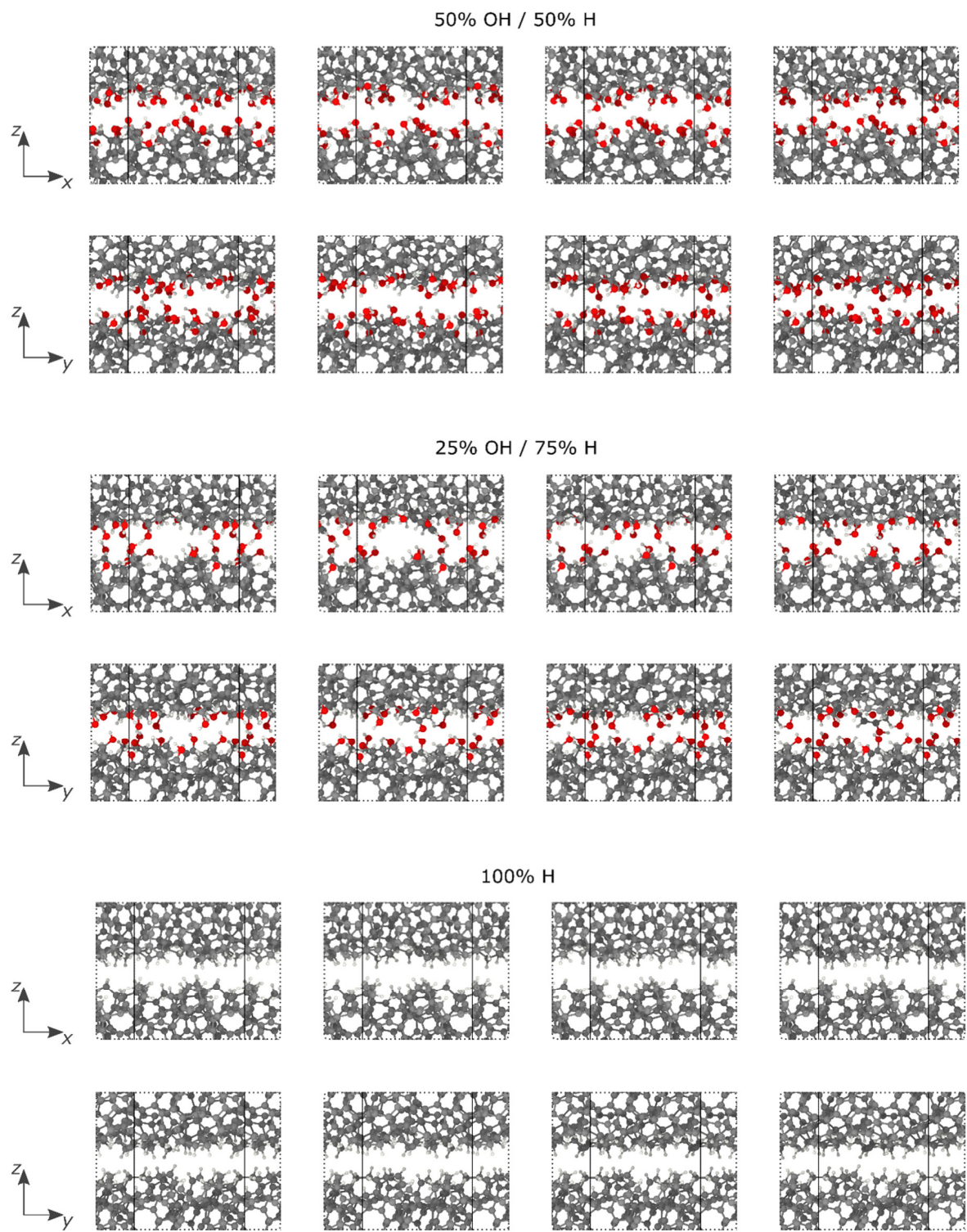

**Fig. S10.** Snapshots of the DFT-relaxed configurations used to calculate adhesion energies for a-C with a bulk density $\rho = 2.0 \text{ g/cm}^3$ (Fig. 11 in the main text). Columns show different lateral positions of the two slabs.

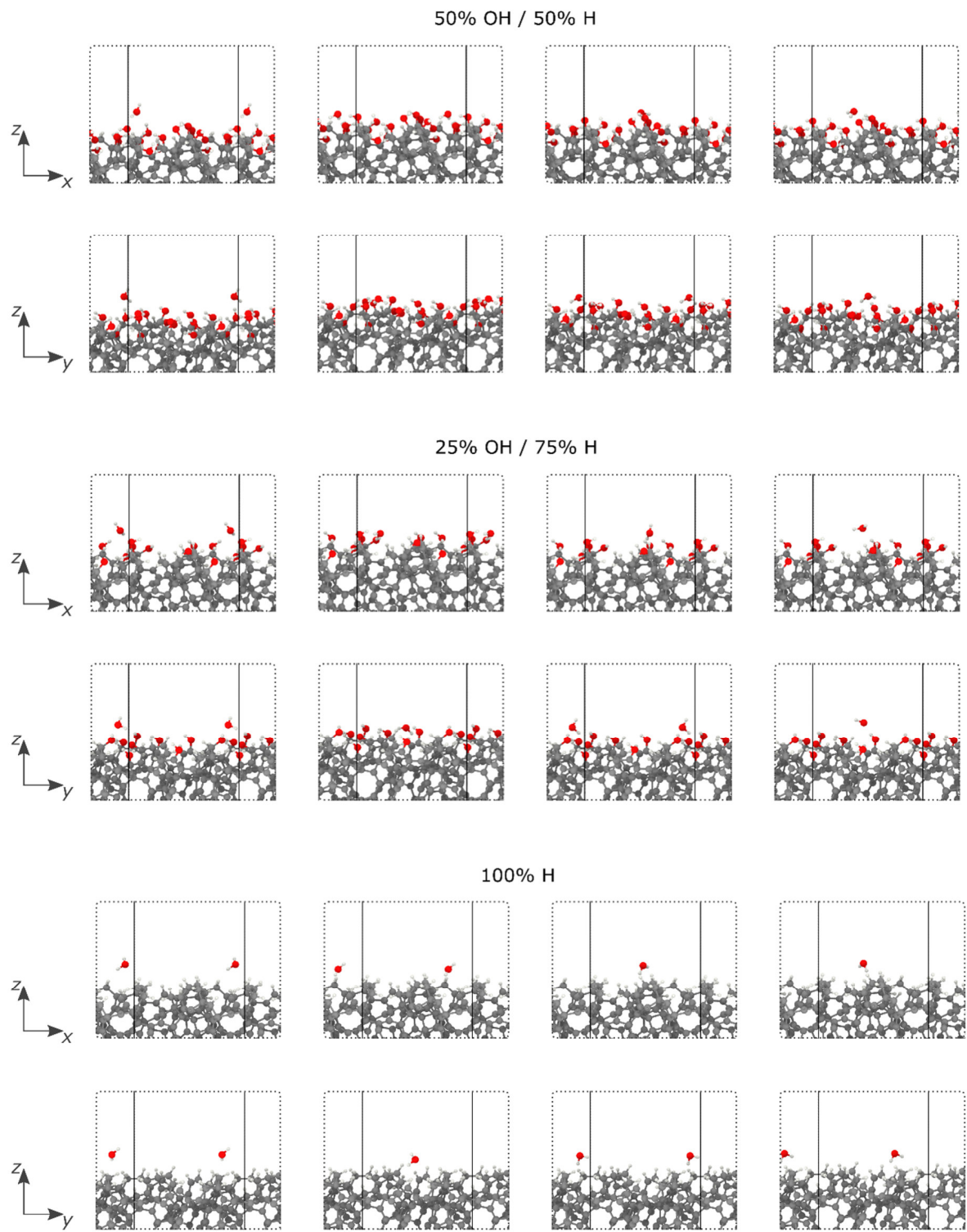

**Fig. S11.** Snapshots of DFT-relaxed configurations used to calculate H$_2$O adsorption energies for a-C with a bulk density $\rho = 2.0$ g/cm$^3$ (Fig. 11 in the main text). The columns show the relaxed configurations corresponding to different lateral positions of the H$_2$O molecule.

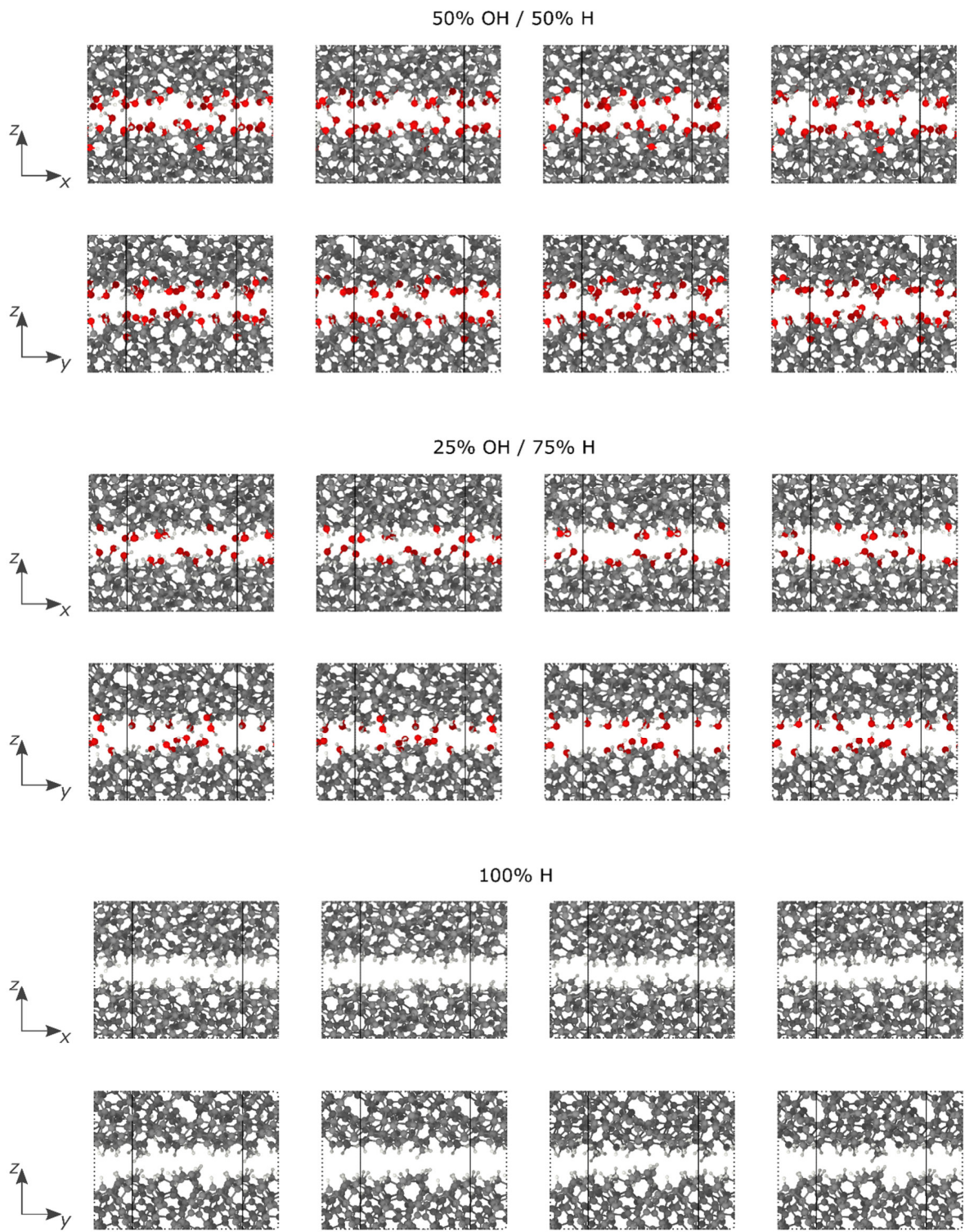

**Fig. S12.** Snapshots of DFT-relaxed configurations used to calculate adhesion energies for a-C with a bulk density $\rho = 2.5$ g/cm$^3$ (Fig. 11 in the main text). Columns show different lateral positions of the two slabs.

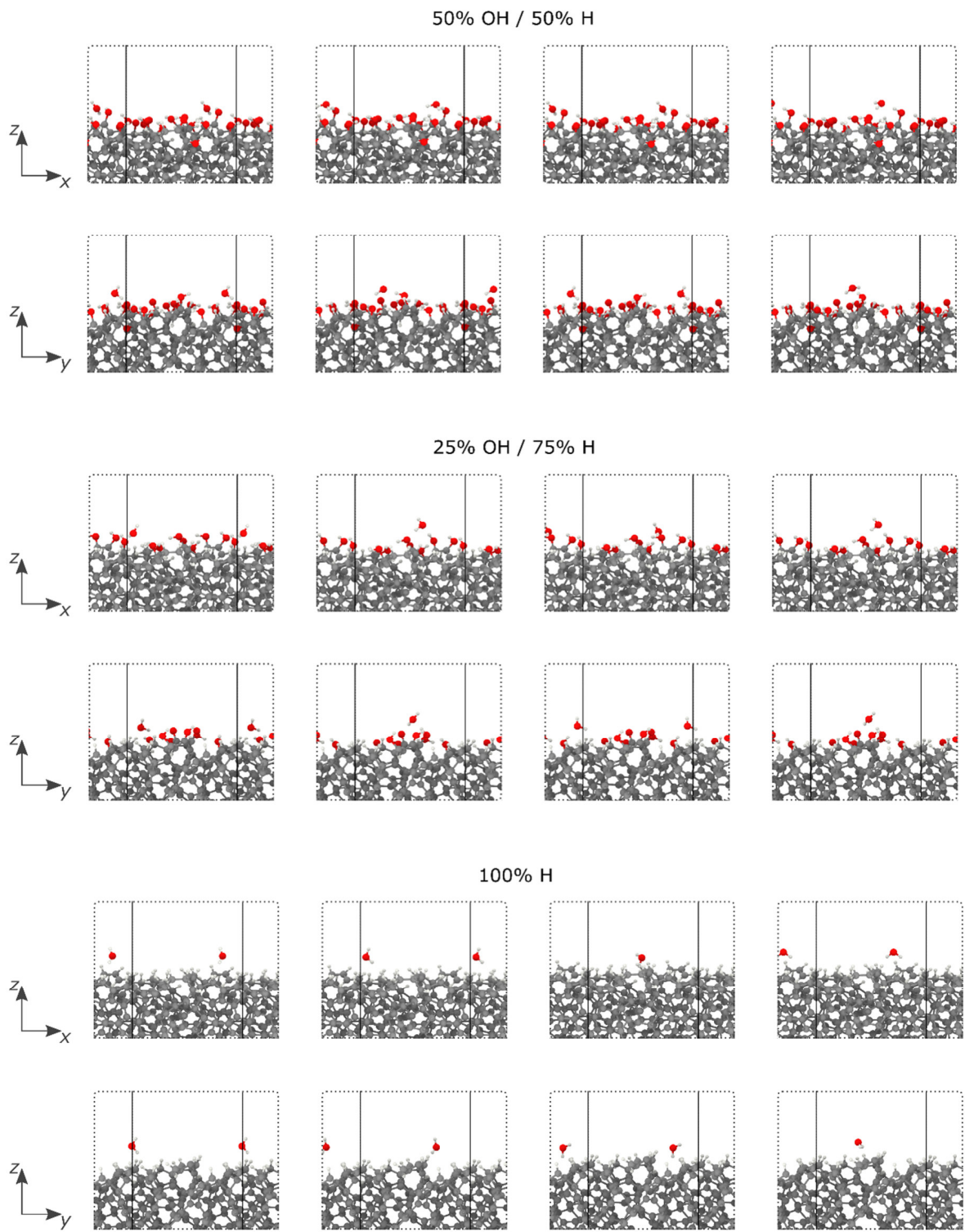

**Fig. S13.** Snapshots of DFT-relaxed configurations used to calculate H$_2$O adsorption energies for a-C with a bulk density $\rho = 2.5$ g/cm$^3$ (Fig. 11 in the main text). The columns show the relaxed configurations corresponding to different lateral positions of the H$_2$O molecule.

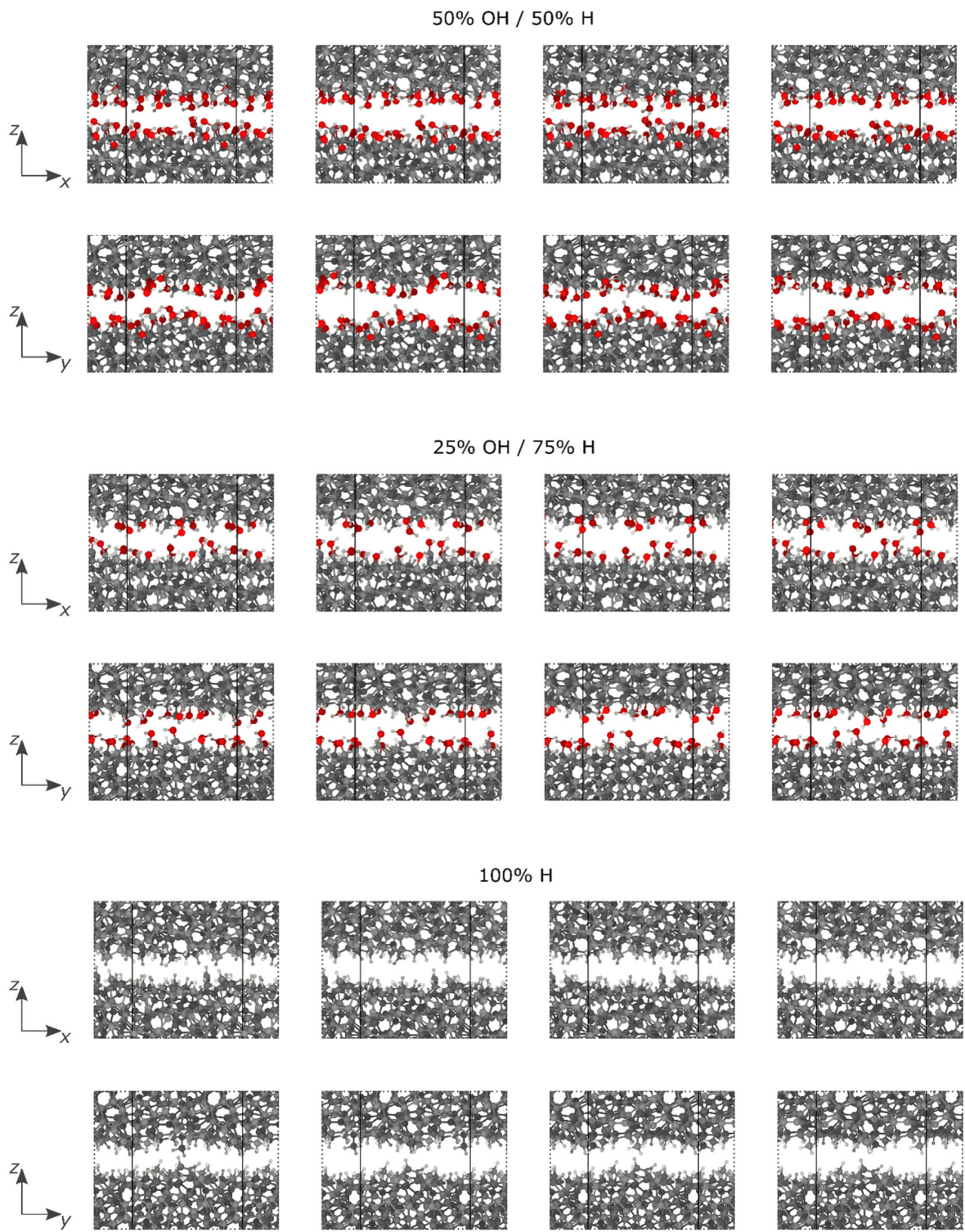

**Fig. S14.** Snapshots of DFT relaxed configurations used to calculate adhesion energies for a-C with a bulk density $\rho = 3.0$ g/cm$^3$ (Fig. 11 in the main text). Columns show different lateral positions of the two slabs.

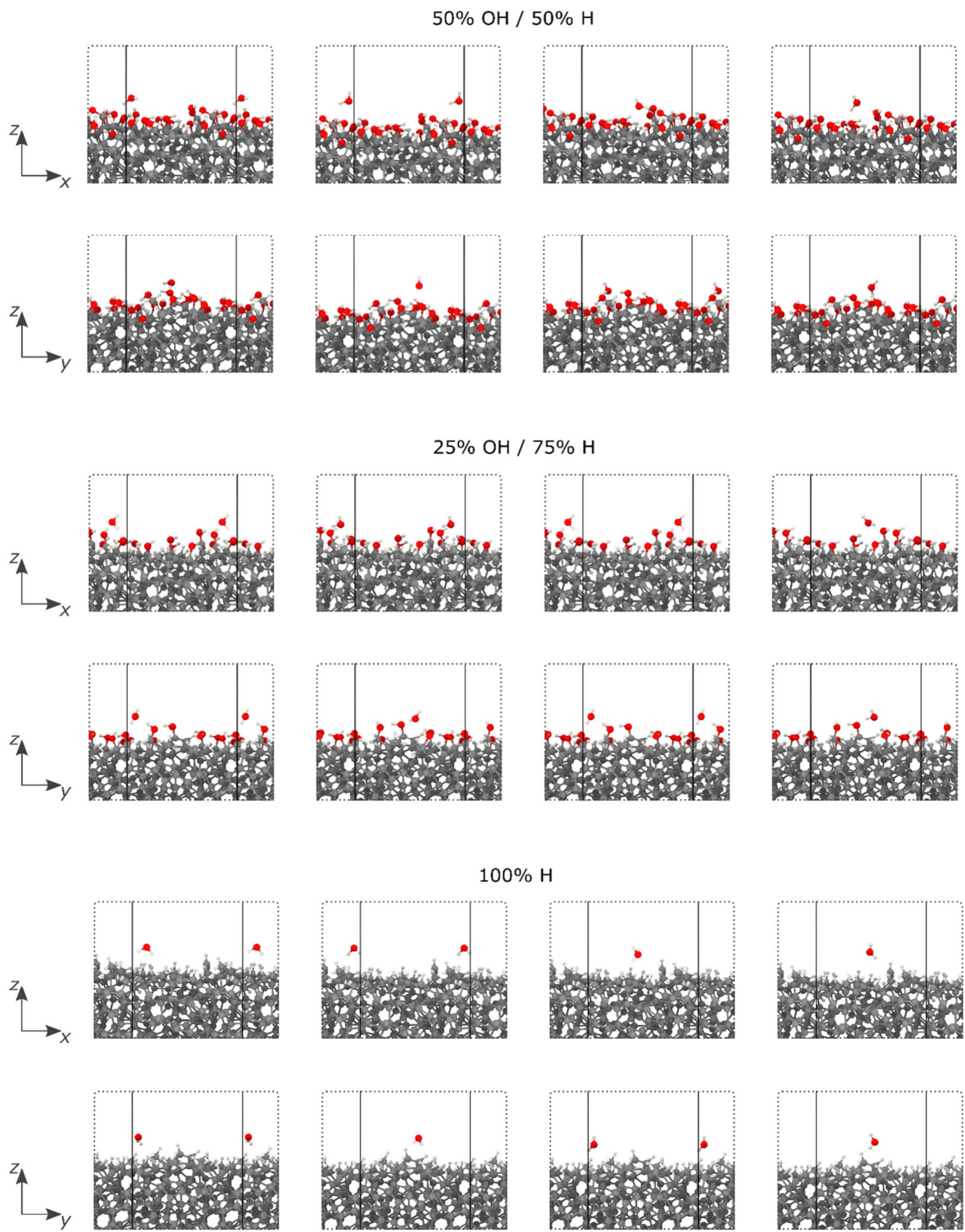

**Fig. S15.** Snapshots of DFT relaxed configurations used to calculate $H_2O$ adsorption energies for a-C bulk density $\rho = 3.0$ g/cm$^3$ (Fig. 11 in the main text). The columns show the relaxed configurations corresponding to different lateral positions of the $H_2O$ molecule.